\documentclass[preprint,12pt]{elsarticle}

\usepackage{amssymb}
\usepackage{amsmath}
\usepackage{hyperref}

\journal{Physica A: Statistical Mechanics and its Applications (accepted manuscript)}

\begin{document}

\begin{frontmatter}

\tnotetext[aam]{This is the author's accepted manuscript of an article published in \emph{Physica A: Statistical Mechanics and its Applications} \textbf{701} (2026) 132041, \href{https://doi.org/10.1016/j.physa.2026.132041}{doi:10.1016/j.physa.2026.132041}. \copyright~2026. This manuscript version is made available under the CC-BY-NC-ND 4.0 license, \url{https://creativecommons.org/licenses/by-nc-nd/4.0/}.}
\title{Stochastic Diversity Stabilization in Nonlinear Competitive
Dynamics: A Statistical Mechanics Approach to Fluctuation
Suppression and Finite-Size Scaling\tnoteref{aam}}

\author[inst1]{Ruocun Ou\corref{cor1}}
\ead{ouruocun@gmail.com}
\cortext[cor1]{Corresponding author.}

\affiliation[inst1]{organization={Nanyang Normal University},
             city={Nanyang},
             postcode={473061},
             state={Henan},
             country={China}}

\begin{abstract}
Stochastic forcing can prevent diversity collapse in competitive
dynamical systems, yet the quantitative conditions and the
dynamical mechanism responsible remain poorly characterized.
We address both questions in a class of nonlinear stochastic
replicator dynamics on the $N$-dimensional probability simplex,
parameterized by a nonlinearity exponent $\gamma$ and additive
Gaussian noise of intensity $\sigma$.
Under the baseline interaction, the minimum noise required
to sustain diversity decays as a power law, $\sigma_c \sim N^{-\alpha}$;
replacing the baseline interaction with a superlinear variant
shifts $\alpha$ from $\approx 1.1$ to $\approx 2.0$,
in agreement with a linear-stability prediction
$\alpha = (\gamma+2)/2$,
identifying $\alpha$ as a model-specific
quantity rather than a universal index.
Scanning the $(\gamma,\sigma)$ plane reveals a smooth
dynamical crossover: the fluctuation susceptibility
$\chi_H = N\,\mathrm{Var}(H)$ peaks at
$\gamma \approx 0.7$ ($\chi_H \approx 0.6$) and
is suppressed at high $\gamma$, with no divergence
near $\gamma \approx 1.4$;
matched finite-size analysis finds no systematic
divergence of the peak with $N$,
ruling out a genuine second-order phase transition.
The entropy relaxation time $\tau_\mathrm{relax}(\gamma)$
remains approximately constant ($\approx 2.9$ time units)
across the full range of $\gamma$, even as $\chi_H$
decreases sixfold; this decoupling between relaxation
timescale and fluctuation amplitude implies that
suppression arises from convergence of diverse pathways
to similar endpoints rather than from accelerated
relaxation along a single channel.
These findings identify \emph{nonlinear fluctuation
suppression via pathway multiplicity}---distinct from
both noise-induced transitions and equilibrium
criticality---as the operative stabilization pattern
in this class of nonequilibrium stochastic systems.
\end{abstract}


\end{frontmatter}


\section{Introduction}
\label{sec:intro}

\subsection{Background}
\label{subsec:background}

In competitive systems comprising multiple interacting species or agents,
deterministic dynamics generically drive the system toward a state of
reduced diversity: one or a few dominant components come to monopolize
the available ``resource'' space while the rest are suppressed.
Such diversity-collapse dynamics arise across a broad class of
nonlinear stochastic systems constrained to a probability simplex,
including species extinction in ecological communities
\cite{ref11,ref12}, fixation in finite-population evolutionary
games \cite{ref14}, winner-take-all convergence in multi-agent
learning and market-share competition, and weight concentration
in neural network ensembles.
Understanding the conditions under which stochastic forcing
can counteract this ordering tendency is therefore a question
of broad relevance to nonlinear science \cite{ref10,ref22}.
This phenomenon, commonly referred to as \emph{diversity collapse} or
competitive exclusion, is a robust prediction of mean-field replicator
dynamics \cite{ref11} and related evolutionary game-theoretic frameworks.
On the $N$-dimensional simplex $\Delta^{N-1}$, the replicator equation
assigns a positive feedback to above-average components and a negative
feedback to below-average ones, creating a self-reinforcing ordering
tendency that terminates in a corner state corresponding to the
extinction of all but a single type.
Such collapse dynamics are well-documented in theoretical ecology,
population genetics, and multi-agent learning, and represent a
fundamental tension with the empirical observation that diversity
is remarkably persistent across biological and social systems.

A natural candidate mechanism for resisting this ordering tendency
is stochasticity.
Random fluctuations, whether arising from finite population size,
environmental variability, or intrinsic demographic noise,
perturb the deterministic flow and can, under appropriate conditions,
prevent the system from becoming trapped at a corner of the simplex.
This stabilizing role of noise has been anticipated on theoretical
grounds in the context of noise-induced transitions \cite{ref10,ref22},
supported by population-level models in evolutionary game theory
\cite{ref11,ref14}, and demonstrated numerically in ecological models
where demographic stochasticity sustains species coexistence that
deterministic dynamics would otherwise eliminate \cite{ref12}.
In the statistical physics language, the question can be reframed
as whether additive stochastic forcing is capable of shifting
the system away from the ordered (low-entropy) absorbing-like
state into a disordered (high-entropy) stationary regime,
in analogy with noise-induced nonequilibrium phase transitions
\cite{ref22}.

Despite this body of work, two important aspects of stochastic
diversity stabilization remain insufficiently understood.
First, the \emph{quantitative parameter dependence} of the
stabilizing effect has not been systematically characterized:
specifically, how the noise--entropy response curve evolves with
system size $N$, what role the deterministic dynamics play at
large $N$, and whether any observed size-dependent pattern is
robust to changes in the microscopic interaction rule.
Second, the \emph{dynamical mechanism} by which stronger nonlinearity
in the competitive interaction suppresses collective fluctuations
has not been identified.
Addressing these two gaps is the primary aim of the present work,
which approaches diversity stabilization as a problem in
nonlinear stochastic dynamics on constrained state spaces
rather than in terms of equilibrium phase transitions.

\subsection{Present work}
\label{subsec:present-work}

To address these questions, we study a class of stochastic replicator-like
systems defined on the $N$-dimensional probability simplex,
in which the competitive interaction is parameterized by a nonlinearity
exponent $\gamma$ and the dynamics are driven by additive Gaussian noise
of strength $\sigma$.
Three main contributions are reported.

\textit{(i) Size-dependent noise--entropy response and deterministic
self-stabilization.}
We systematically measure the noise--entropy response curve
$\langle H \rangle(\sigma)$ across system sizes
$N \in \{5, 8, 10, 15, 20, 30, 50\}$ and reveal a
qualitative transition in stabilization mechanism:
for small $N$ the entropy at $\sigma = 0$ is well below the
high-diversity regime, so external noise is the primary
stabilizing agent; for large $N$ the deterministic dynamics
alone sustain a high-entropy stationary state, and noise plays
only a secondary modulating role.
In the small-$N$ regime ($N \leq 15$) the noise intensity
at the steepest response point decreases approximately as
$\sigma_c \sim N^{-\alpha}$; however, this approximate
scaling breaks down at large $N$, and the exponent
$\alpha$ shifts from $\approx 1$ to $\approx 2$ as the
nonlinearity $\gamma$ increases, in agreement with the
analytical prediction $\alpha = (\gamma+2)/2$
(Sections~\ref{subsec:fss-scaling}--\ref{subsec:universality},
\ref{app:bound}),
confirming that it reflects the specific balance of the
baseline replicator interaction rather than a universal law.

\textit{(ii) Nonlinear phase structure and fluctuation analysis.}
We construct a dynamical phase diagram in the $(\gamma, \sigma)$
parameter plane and show that the system undergoes a smooth crossover
rather than a sharp phase transition as $\gamma$ increases.
To test for criticality, we compute the fluctuation susceptibility
$\chi_H = N\,\mathrm{Var}(H)$, where $H$ is the normalized Shannon
entropy, across the full parameter space.
The susceptibility peaks at $\gamma \approx 0.7$
($\chi_H \approx 0.6$) and is suppressed at high $\gamma$;
no divergence is observed near $\gamma \approx 1.4$,
where a naive visual inspection of the phase diagram might suggest
a structural change.
Matched finite-size analysis at $N \in \{10, 20, 40\}$ finds
no systematic divergence of the peak susceptibility with $N$,
in direct contradiction to expectations near a
genuine second-order phase transition \cite{ref4,ref5}.
These findings establish that the high-$\gamma$ regime is not a
critical state but an \emph{entropy-saturated nonlinear phase}
in which collective fluctuations are suppressed by the nonlinear
interaction rather than amplified.

\textit{(iii) Relaxation dynamics and mechanistic interpretation.}
We measure the macroscopic entropy relaxation time $\tau_\mathrm{relax}$
as a function of $\gamma$ using a stationary-distribution criterion
appropriate for noisy systems \cite{ref7,ref8}.
The resulting $\tau_\mathrm{relax}(\gamma)$ curve is approximately
flat, with mean values between $2.5$ and $3.2$ time units across
the full $\gamma$ range and no systematic dependence on
nonlinearity strength.
This insensitivity of the relaxation timescale stands in sharp
contrast to the sixfold decrease of $\chi_H$ over the same
$\gamma$ range, revealing a decoupling between the rate at which
individual trajectories reach stationarity and the degree of
ensemble-level fluctuation suppression.
The decoupling implies that strong nonlinear coupling suppresses
collective fluctuations not by accelerating convergence toward
a single deep attractor, but by routing different trajectories
through diverse microscopic pathways that nonetheless converge
to similar macroscopic endpoints---a mechanism we term
\emph{pathway-multiplicity-driven ensemble decorrelation}.

The remainder of the paper is organized as follows.
Section~\ref{sec:model} defines the model and all observables.
Sections~\ref{sec:results-scaling}--\ref{sec:results-relaxation}
present the three groups of numerical results.
Section~\ref{sec:discussion} discusses the findings in relation
to existing theoretical frameworks, and Section~\ref{sec:conclusion}
summarizes the conclusions.

\section{Model and Methods}
\label{sec:model}

\subsection{Stochastic dynamics on the simplex}
\label{subsec:model}

We consider a system of $N$ interacting components whose states are
represented by the vector $\mathbf{x} = (x_1, x_2, \ldots, x_N)$
constrained to the $(N{-}1)$-dimensional probability simplex,
\begin{equation}
    \Delta^{N-1} = \left\{ \mathbf{x} \in \mathbb{R}^N \;\middle|\;
    x_i \geq 0 \;\forall\, i,\quad \sum_{i=1}^{N} x_i = 1 \right\}.
    \label{eq:simplex}
\end{equation}
Each component $x_i$ can be interpreted as the relative abundance
of type $i$ in a population, the market share of agent $i$,
or more generally the probability weight assigned to option $i$
in a finite mixture.
The simplex geometry ensures that competition is inherently
zero-sum: an increase in $x_i$ must be offset by a decrease
elsewhere.

\subsubsection*{Baseline interaction}

The deterministic driving force is taken from the class of
replicator-like dynamics \cite{ref11}.
In the baseline model, the drift term for component $i$ is
\begin{equation}
    f_i(\mathbf{x}) = x_i \bigl(x_i - \langle x \rangle\bigr),
    \qquad \langle x \rangle = \frac{1}{N}\sum_{j=1}^{N} x_j
    = \frac{1}{N},
    \label{eq:baseline}
\end{equation}
where the equality $\langle x \rangle = 1/N$ follows from the
simplex constraint.
The sign structure of $f_i$ is that of a positive feedback:
components above the population mean gain a positive drift and
those below the mean are driven further downward.
Under purely deterministic dynamics ($\sigma = 0$), this creates
a self-reinforcing ordering that terminates at a vertex of the
simplex, corresponding to the extinction of all but one type---the
diversity-collapse state described in Section~\ref{subsec:background}.

\subsubsection*{Generalized nonlinear interaction}

To study how the strength of the competitive nonlinearity
affects the collective behavior, we introduce the
one-parameter family
\begin{equation}
    f_i(\mathbf{x};\gamma) = x_i
    \Bigl( x_i^{\,\gamma} - \langle x^{\gamma} \rangle \Bigr),
    \qquad
    \langle x^{\gamma} \rangle = \frac{1}{N}\sum_{j=1}^{N} x_j^{\,\gamma},
    \label{eq:generalized}
\end{equation}
where $\gamma > 0$ is the nonlinearity exponent.
The baseline model~\eqref{eq:baseline} is recovered at $\gamma = 1$.
For $\gamma > 1$ the positive feedback is superlinear:
high-abundance components receive a disproportionately amplified
advantage, making competitive exclusion mechanistically stronger.
The exponent $\gamma$ therefore serves as the primary control
parameter governing the type and intensity of collective behavior
investigated in Sections~\ref{sec:results-phase}
and~\ref{sec:results-relaxation}.

\subsubsection*{Linear stability of the uniform fixed point}

The uniform state $x_i = 1/N$ for all $i$ is a fixed point of
the deterministic dynamics~\eqref{eq:generalized} for any
$\gamma > 0$.
Linearizing around this point, the Jacobian of the drift
restricted to the simplex tangent space takes the form
\begin{equation}
    J_{ij}\big|_{1/N} = \gamma\,N^{-\gamma}
    \!\left(\delta_{ij} - \frac{1}{N}\right)
    = \gamma\,N^{-\gamma}\,P_{ij},
    \label{eq:jacobian}
\end{equation}
where $P = I - N^{-1}\mathbf{1}\mathbf{1}^\top$ is the
orthogonal projector onto the zero-sum subspace
(the tangent space of $\Delta^{N-1}$).
On this subspace $P$ acts as the identity, so $J$ has a
single eigenvalue
\begin{equation}
    \lambda(\gamma,N) = \gamma\,N^{-\gamma}
    \label{eq:eigenvalue}
\end{equation}
with multiplicity $N-1$.
Since $\lambda > 0$, the uniform state is linearly
\emph{unstable}: small perturbations grow at rate $\lambda$,
driving the system toward the simplex vertices (diversity
collapse).

Two features of~\eqref{eq:eigenvalue} connect directly to
the numerical results reported below:
(i)~at fixed $N$, the instability rate decreases rapidly with
$\gamma$ (e.g., from $0.05$ at $\gamma=1$ to $0.005$ at
$\gamma=2$ for $N=20$), providing a first-principles
explanation for the self-stabilization of the high-entropy
state at high $\gamma$ (Section~\ref{sec:results-phase});
(ii)~at fixed $\gamma$, $\lambda \to 0$ as $N \to \infty$,
explaining why large systems self-stabilize without noise
(Section~\ref{subsec:fss-scaling}).
The generalized analytical bound $\sigma_c(N,\gamma)$
derived from~\eqref{eq:eigenvalue} is presented in
\ref{app:bound}.

\subsubsection*{Stochastic differential equation and numerical integration}

Additive Gaussian noise of uniform intensity $\sigma \geq 0$
is superimposed on the deterministic drift,
yielding the It\^{o} stochastic differential equation
\begin{equation}
    \mathrm{d}x_i = f_i(\mathbf{x};\gamma)\,\mathrm{d}t
    + \sigma\,\mathrm{d}W_i(t),
    \qquad i = 1, \ldots, N,
    \label{eq:sde}
\end{equation}
where $W_i(t)$ are independent standard Wiener processes \cite{ref7,ref8}.
Equation~\eqref{eq:sde} is written in the It\^{o} sense;
the simplex projection described below introduces an implicit
reflecting boundary that maintains the probabilistic interpretation
of $\mathbf{x}$ throughout the simulation.

Following the standard formulation of stochastic processes
on bounded state spaces \cite{ref7,ref8}, equation~\eqref{eq:sde}
is integrated numerically using the
Euler--Maruyama scheme \cite{ref17,ref20},
\begin{equation}
    x_i(t + \Delta t) = x_i(t)
    + f_i\bigl(\mathbf{x}(t);\gamma\bigr)\,\Delta t
    + \sigma\,\sqrt{\Delta t}\;\xi_i,
    \qquad \xi_i \overset{\mathrm{i.i.d.}}{\sim} \mathcal{N}(0,1),
    \label{eq:em}
\end{equation}
with time step $\Delta t = 10^{-3}$.
After each update, simplex membership is enforced by a two-step
projection: (i) all negative entries are clipped to a small
positive floor $\varepsilon = 10^{-10}$, and (ii) the vector
is renormalized to unit sum,
\begin{equation}
    x_i \;\leftarrow\; \frac{\max(x_i,\,\varepsilon)}
    {\displaystyle\sum_{j=1}^{N}\max(x_j,\,\varepsilon)}.
    \label{eq:projection}
\end{equation}
This projection preserves the simplex constraint at every step
without modifying the qualitative dynamics, as verified by
the insensitivity of all reported results to the exact value
of $\varepsilon$ over several orders of magnitude.
Constrained SDEs on the simplex admit
several well-posed numerical treatments:
elimination of one component via the constraint
$x_N = 1 - \sum_{i=1}^{N-1} x_i$ and integration in
reduced coordinates (e.g.\ the log-ratio coordinates
used by Imhof \cite{ref26} for stochastic replicator
dynamics on $\Delta^{N-1}$), and step-wise projection
back onto the feasible set after each Euler--Maruyama
update (e.g.\ Bubeck, Eldan \& Lehec \cite{ref25},
who show that projected Langevin schemes discretize
the corresponding reflected diffusion in the sense of
Lions \& Sznitman \cite{ref24}).
In the simplex interior these approaches integrate
the same reduced dynamics;
we adopt the projection scheme~\eqref{eq:projection}
because it treats all $N$ components symmetrically
and handles the boundary uniformly via the floor
$\varepsilon$.
The floor also prevents the $x_i \ln x_i \to 0$ term from
producing numerical underflow in the entropy calculation;
since $\varepsilon = 10^{-10}$ contributes at most
$\varepsilon \ln \varepsilon \approx 2.3 \times 10^{-9}$
per component to the raw entropy sum, the systematic bias
on $H$ is negligible at all $\sigma$ values explored.

\subsection{Observables}
\label{subsec:observables}

\subsubsection*{Normalized Shannon entropy}

The diversity of the system state $\mathbf{x}$ is quantified by
the normalized Shannon entropy
\begin{equation}
    H(\mathbf{x}) = -\frac{1}{\ln N}
    \sum_{i=1}^{N} x_i \ln x_i,
    \label{eq:entropy}
\end{equation}
where the prefactor $(\ln N)^{-1}$ maps the range of $H$ onto
the unit interval: $H = 0$ corresponds to a fully ordered
(single-dominant) corner state, and $H = 1$ to the uniform
distribution $x_i = 1/N$ for all $i$.
The normalization is essential when comparing results across
system sizes, because the raw Shannon entropy $-\sum x_i \ln x_i$
grows with $N$ at fixed diversity level, introducing a spurious
size dependence that would systematically bias estimates of the
critical noise intensity $\sigma_c$ toward smaller values
for larger $N$.
All entropy values reported in this paper refer to the
normalized form~\eqref{eq:entropy}.

\subsubsection*{Critical noise intensity}

For a given system size $N$ and nonlinearity exponent $\gamma$,
the critical noise intensity $\sigma_c$ is defined as the
smallest value of $\sigma$ for which the long-time average of
$H$ exceeds a prescribed diversity threshold $H_\mathrm{thr}$,
\begin{equation}
    \sigma_c(N,\gamma) = \min\bigl\{
    \sigma \geq 0 \;:\; \langle H \rangle_\mathrm{ss} > H_\mathrm{thr}
    \bigr\},
    \label{eq:sigma_c}
\end{equation}
where $\langle \cdot \rangle_\mathrm{ss}$ denotes a steady-state
time average taken after a burn-in period of at least
$3\times 10^3$ integration steps.
We set $H_\mathrm{thr} = 0.7$ throughout; results are qualitatively
insensitive to moderate variations of this threshold.
The quantity $\langle H \rangle_\mathrm{ss}$ is a deterministic
quantity obtained by averaging over both time (post-burn-in trajectory)
and independent trials; the threshold condition therefore involves
no ambiguity from instantaneous stochastic fluctuations.
The quantity $\sigma_c$ measures the minimum stochastic forcing
needed to sustain a diverse stationary state against the
ordering tendency of the deterministic drift.

\subsubsection*{Fluctuation susceptibility}

To probe collective fluctuations and test for critical behavior,
we define the fluctuation susceptibility
\begin{equation}
    \chi_H = N\,\mathrm{Var}(H),
    \label{eq:chi}
\end{equation}
in direct analogy with the magnetic susceptibility
$\chi \propto N\,\mathrm{Var}(m)$ used in finite-size scaling
analysis of spin systems \cite{ref4,ref5}.
Here $\mathrm{Var}(H)$ is the variance of $H(\mathbf{x}(t))$
computed over the steady-state trajectory.
Physically, a large $\chi_H$ indicates that independent
realizations of the dynamics produce widely different
entropy levels---the system is sensitive to initial conditions
and has not settled into a unique macroscopic state.
A small $\chi_H$ indicates that all realizations converge
to a similar entropy, signaling robust self-organization.
The factor of $N$ rescales the susceptibility so that a
genuine second-order phase transition would manifest as a
divergence $\chi_H \sim N^{\eta/\nu}$ in the thermodynamic
limit, where $\eta$ and $\nu$ are standard critical exponents,
while a smooth crossover or finite-size artefact would
produce a $\chi_H$ that either saturates or decreases with
increasing $N$ \cite{ref4,ref5}.
This distinction is exploited in Section~\ref{sec:results-phase}
to determine whether the structural changes observed in the
phase diagram correspond to true critical behavior.

\subsubsection*{Entropy relaxation time}

The relaxation time $\tau_\mathrm{relax}$ measures how quickly
the macroscopic entropy $H(t)$ reaches its stationary distribution
from a random initial condition drawn uniformly on $\Delta^{N-1}$.

A natural but physically incorrect approach would be to declare
relaxation complete once the microscopic state vector ceases
to change appreciably, e.g., when
$\frac{1}{N}\sum_i |x_i(t) - x_i(t-\Delta t)| < \varepsilon_\mathrm{micro}$.
In a noise-driven system, however, individual components $x_i$
are permanently agitated by the Wiener increments in~\eqref{eq:sde}
and never reach a microscopic fixed point \cite{ref7,ref8}.
Applying such a criterion causes the simulation to run until
the maximum allowed time in all parameter regimes, producing
a spurious flat $\tau_\mathrm{relax}(\gamma)$ curve that carries
no physical information (see~\ref{app:criterion}).

We therefore adopt a \emph{macroscopic} convergence criterion
based on the stability of the running entropy average.
The simulation trajectory is partitioned into non-overlapping
windows of $n_w = 100$ integration steps.
At each window boundary we compute the window-averaged entropy
$\bar{H}_k$ for the current window $k$ and the preceding window
$k-1$.
Relaxation is declared when the absolute difference
$|\bar{H}_k - \bar{H}_{k-1}|$ remains below a tolerance
$\varepsilon_H = 10^{-4}$ for $n_s = 5$ consecutive window pairs,
\begin{equation}
    \tau_\mathrm{relax} = \Delta t \cdot t^*,
    \qquad
    t^* = \min\!\left\{ t \;:\;
    \bigl|\bar{H}_{k}(t) - \bar{H}_{k-1}(t)\bigr| < \varepsilon_H
    \;\text{ for } n_s \text{ consecutive } k
    \right\}.
    \label{eq:tau}
\end{equation}
This criterion is immune to microscopic noise fluctuations
because $H(\mathbf{x})$ is a smooth, $N$-fold averaged function
of the state: its running mean stabilizes as soon as the
macroscopic distribution reaches stationarity,
regardless of the ongoing microscopic agitation \cite{ref7,ref8}.
The choices $n_w = 100$, $\varepsilon_H = 10^{-4}$, and
$n_s = 5$ were verified to produce qualitatively identical
$\tau_\mathrm{relax}(\gamma)$ curves when varied by factors
of two in either direction; detailed sensitivity checks
are provided in Supplementary Material~S5.

\subsection{Simulation parameters}
\label{subsec:params}

All numerical experiments share a common baseline configuration
summarized in Table~\ref{tab:params}.
The system size $N$ and the nonlinearity exponent $\gamma$ are
varied systematically as described below for each group of
experiments; all other parameters are held fixed.

\begin{table}[htbp]
\centering
\caption{Baseline simulation parameters used throughout this work.
Parameters varied in specific experiments are stated in the
corresponding results sections.}
\label{tab:params}
\begin{tabular}{lll}
\hline
Parameter & Symbol & Value \\
\hline
Integration time step       & $\Delta t$          & $10^{-3}$ \\
Total simulation steps      & $T$                 & $6\times 10^{3}$ \\
Burn-in steps               & $T_\mathrm{burn}$   & $3\times 10^{3}$ \\
Window size (relaxation)    & $n_w$               & $100$ steps \\
Stable windows required     & $n_s$               & $5$ \\
Entropy convergence tol.    & $\varepsilon_H$     & $10^{-4}$ \\
Simplex floor               & $\varepsilon$       & $10^{-10}$ \\
Diversity threshold         & $H_\mathrm{thr}$   & $0.7$ \\
Independent trials          & ---                 & $\geq 5$ (up to 20) \\
\hline
\end{tabular}

\smallskip
\noindent\textit{Note.} The relaxation analysis
(Section~\ref{sec:results-relaxation}) uses
$\Delta t = 0.01$ and $T = 6000$ steps (total simulation
time $60$ units) to ensure macroscopic convergence;
all other experiments use the baseline $\Delta t = 10^{-3}$.
The window parameters ($n_w$, $n_s$, $\varepsilon_H$)
are defined in integration steps, so their physical
durations scale with $\Delta t$: e.g., $n_w = 100$ steps
corresponds to $0.1$ time units in the baseline experiments
and $1.0$ time unit in the relaxation analysis.
\end{table}

\paragraph{Critical noise scaling (Section~\ref{sec:results-scaling}).}
The system size is varied over $N \in \{5, 8, 10, 15, 20\}$ and
the noise intensity is scanned over a uniform grid of
$\sigma \in [0,\, 0.20]$ with step $\Delta\sigma = 0.005$.
For each $(N, \sigma)$ pair, five independent trials are run
from distinct random initial conditions drawn uniformly on
$\Delta^{N-1}$; the reported $\langle H \rangle_\mathrm{ss}$
is the mean over trials and over the post-burn-in trajectory.
Note that $N = 20$ lies in the deterministically self-stabilized
regime (Section~\ref{subsec:fss-scaling}) and is therefore
excluded from the power-law fit; only
$N \in \{5, 8, 10, 15\}$ are used for the scaling analysis.
The universality test uses the same grid with the interaction
replaced by the superlinear form~\eqref{eq:generalized} at
$\gamma \in \{1.0, 1.2, 1.4, 1.6, 1.8, 2.0\}$.

\paragraph{Phase diagram and fluctuation analysis
(Section~\ref{sec:results-phase}).}
The nonlinearity exponent and noise intensity are scanned jointly
over $\gamma \in [0.5,\, 2.0]$ (14 uniformly spaced values) and
$\sigma \in [0,\, 0.08]$ (35 uniformly spaced values),
with $N = 20$ fixed.
Twenty independent trials are run per parameter pair;
the fluctuation susceptibility $\chi_H$ is computed from the
pooled post-burn-in entropy time series across all trials.
For the finite-size robustness check, the same $(\gamma, \sigma)$
grid is repeated at $N \in \{10, 20, 40\}$, also with
twenty trials per point.
Numerical stability with respect to the integration step is
verified by repeating the $N = 20$ grid at
$\Delta t = 5 \times 10^{-4}$.

\paragraph{Relaxation dynamics (Section~\ref{sec:results-relaxation}).}
The nonlinearity exponent is varied over $\gamma \in [0.25,\, 2.0]$
(15 uniformly spaced values) at fixed $N = 20$ and
$\sigma = 0.03$, a value within the range where the
susceptibility peak is observed (Section~\ref{sec:results-phase}).
For these runs a coarser time step $\Delta t = 0.01$ is used
to achieve a total simulation time of $60$ time units
(6000 steps), ensuring that the macroscopic entropy has
ample time to reach stationarity; the coarser step is
justified by the convergence check in Supplementary~S5.
Twenty independent trials are run per $\gamma$ value to obtain
reliable estimates of the mean and standard error of
$\tau_\mathrm{relax}$; raw trial-level distributions are also
reported to characterize the trial-to-trial variability.

\paragraph{Numerical validation summary.}
The robustness of all reported results has been verified through
four independent checks, which are documented in the
Supplementary Materials:
(i) integration time-step convergence
($\Delta t = 10^{-3}$ vs $5\times10^{-4}$, Supplementary~S5);
(ii) finite-size scaling across $N \in \{10, 20, 40\}$
(Supplementary~S6);
(iii) sensitivity of $\tau_\mathrm{relax}$ to the window
parameters $n_w$ and $\varepsilon_H$, verified to produce
qualitatively identical curves when varied by factors of two;
and (iv) insensitivity of the simplex-floor parameter
$\varepsilon = 10^{-10}$ over several orders of magnitude,
with the resulting entropy bias bounded at $< 10^{-8}$.
Together, these checks confirm that the conclusions of
Sections~\ref{sec:results-scaling}--\ref{sec:results-relaxation}
are not artefacts of specific numerical choices.

\section{Noise-induced stabilization and size-dependent diversity response}
\label{sec:results-scaling}

\subsection{Basic phenomenology: noise arrests diversity collapse}
\label{subsec:phenomenology}

\begin{figure}[htbp]
    \centering
    \includegraphics[width=\linewidth]{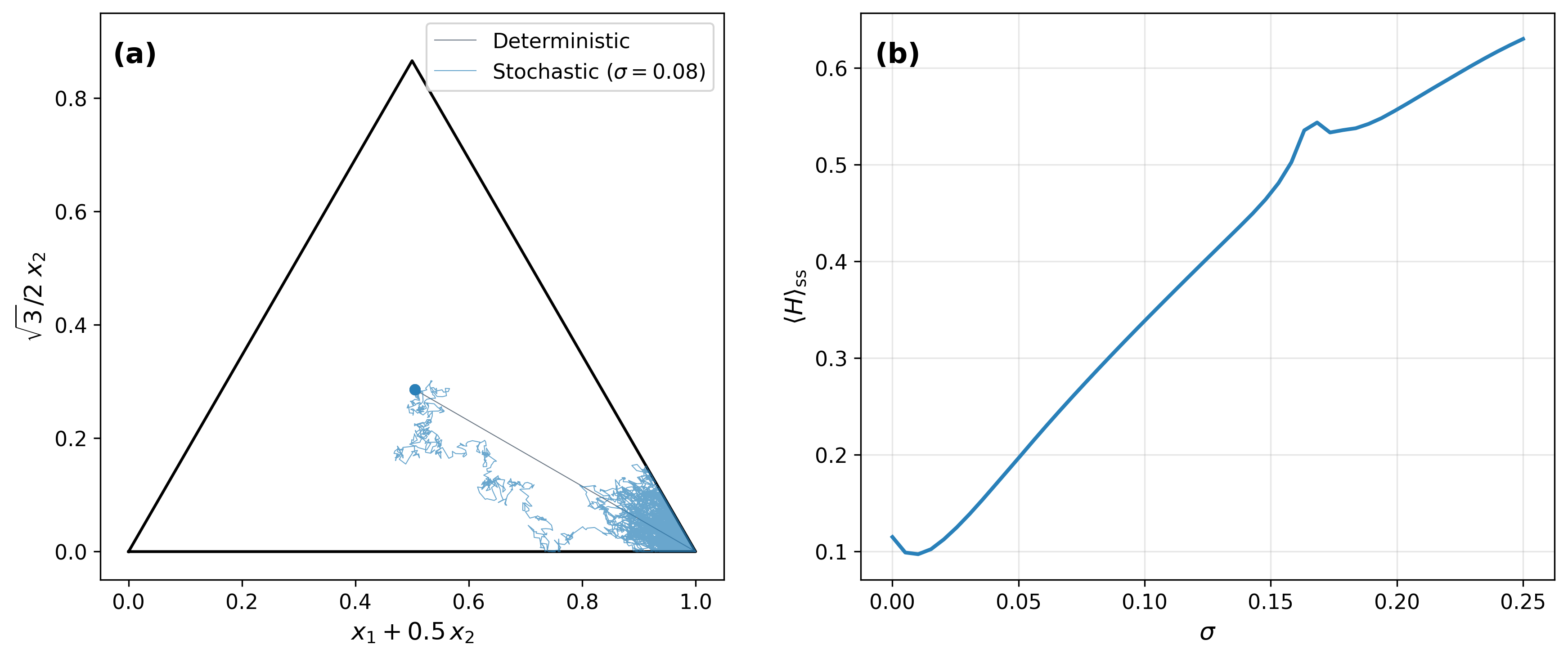}
    \caption{Basic phenomenology of stochastic diversity stabilization
    ($N = 3$, $\gamma = 1$).
    \textbf{(a)} Representative trajectories on the two-dimensional simplex
    for deterministic dynamics ($\sigma = 0$, left) and stochastic dynamics
    ($\sigma = 0.08$, right).
    The deterministic trajectory collapses to a vertex within $\sim\!2\times10^3$
    steps; the stochastic trajectory explores the simplex interior
    throughout $5\times10^3$ steps.
    \textbf{(b)} Steady-state mean normalized entropy
    $\langle H \rangle_\mathrm{ss}$ as a function of noise strength $\sigma$.
    The monotonic increase confirms that noise systematically promotes
    diversity; saturation near $\langle H \rangle_\mathrm{ss} \approx 0.73$
    reflects the geometric ceiling of the $N = 3$ simplex.}
    \label{fig:basic}
\end{figure}

We begin by establishing the qualitative contrast between deterministic
and stochastic dynamics on the simplex at $N = 3$, $\gamma = 1$.
Figure~\ref{fig:basic}a shows representative trajectories in the
two-dimensional simplex for a deterministic run ($\sigma = 0$)
and a stochastic run ($\sigma = 0.08$), both started from
the near-uniform initial condition $\mathbf{x}_0 \approx (1/3, 1/3, 1/3)$.
Under deterministic dynamics the system follows a direct path toward
a vertex, reaching the diversity-collapse state within approximately
$2\times10^3$ integration steps.
Under stochastic dynamics the trajectory explores the interior of
the simplex throughout the full simulation window of
$5\times10^3$ steps, with no discernible drift toward any vertex.

The dependence of long-time diversity on noise intensity is
quantified in Figure~\ref{fig:basic}b, which shows the
steady-state mean entropy $\langle H \rangle_\mathrm{ss}$
as a function of $\sigma$ for $N = 3$, $\gamma = 1$.
The curve rises monotonically from $\langle H \rangle_\mathrm{ss} \approx 0$
at $\sigma = 0$ and saturates near $\langle H \rangle_\mathrm{ss} \approx 0.73$
for $\sigma \gtrsim 0.20$.
The monotonic character confirms that noise acts as a
diversity-promoting force throughout the range explored,
consistent with the general theory of noise-induced
stabilization \cite{ref10,ref22}.
The saturation reflects the finite-entropy ceiling imposed by
the $N = 3$ simplex geometry rather than any dynamical transition.

\subsection{Size-dependent noise--entropy response}
\label{subsec:fss-scaling}

To characterize how the role of noise changes with system size,
we compute the full noise--entropy response curve
$\langle H \rangle_\mathrm{ss}(\sigma)$ for each
$N \in \{5, 8, 10, 15, 20, 30, 50\}$ under the
baseline interaction ($\gamma = 1$).

\begin{figure}[htbp]
    \centering
    \includegraphics[width=0.75\linewidth]{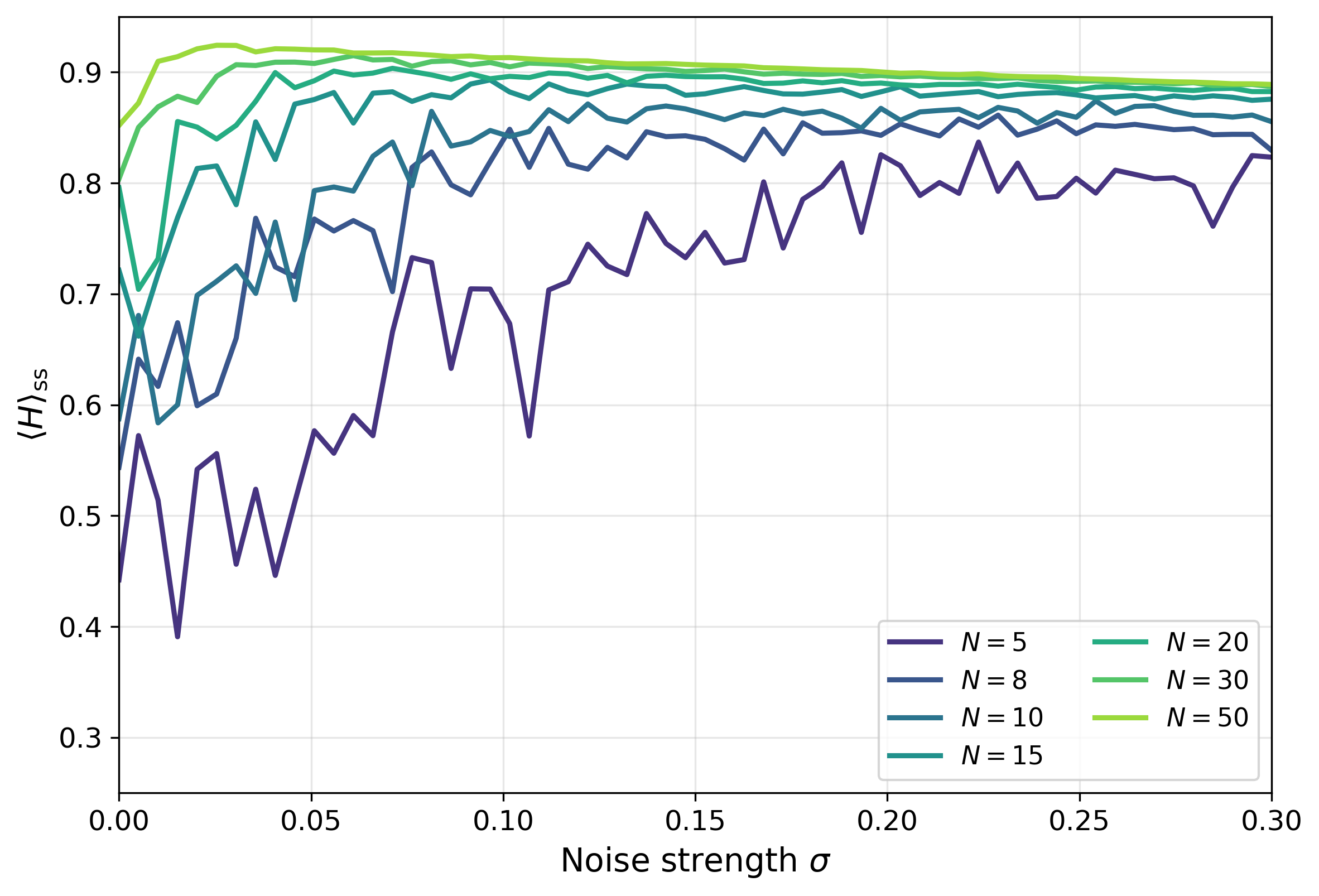}
    \caption{Noise--entropy response curves
    $\langle H \rangle_\mathrm{ss}(\sigma)$ for system sizes
    $N \in \{5, 8, 10, 15, 20, 30, 50\}$ under the baseline
    interaction ($\gamma = 1$; 10 trials per point).
    Dashed vertical lines mark the inflection point $\sigma_c$
    (location of steepest response) for each $N$.
    Small systems ($N \leq 15$, dark tones) show a pronounced
    noise-driven entropy rise from a low baseline, indicating
    that external noise is the primary stabilizing agent.
    Large systems ($N \geq 20$, light tones) already maintain
    high entropy at $\sigma = 0$; their response to noise is
    shallow, reflecting deterministic self-stabilization.
    The crossover between these two regimes is the central
    result of this section.}
    \label{fig:scaling}
\end{figure}

Figure~\ref{fig:scaling} shows all seven response curves together.
Two qualitatively distinct behaviors are visible.
For small $N$ ($N \leq 15$), the entropy at zero noise is
substantially below the high-diversity region
($\langle H \rangle|_{\sigma=0} \approx 0.44$--$0.72$),
and increasing $\sigma$ produces a pronounced, steep rise
in $\langle H \rangle$: noise is the primary agent driving
the system away from the ordering tendency of the
deterministic drift.
For large $N$ ($N \geq 20$), the entropy at $\sigma = 0$
is already elevated ($\langle H \rangle|_{\sigma=0} \approx
0.80$--$0.85$), and the response to increasing $\sigma$ is
shallow: the deterministic dynamics alone sustain a
high-diversity stationary state, and noise plays only a
secondary modulating role.
This crossover in stabilization mechanism---from
noise-dominated at small $N$ to deterministically
self-stabilized at large $N$---is the central finding
of this section and connects to the large-$N$ self-organization
discussed in evolutionary game dynamics \cite{ref14}.

Within the noise-dominated small-$N$ regime, the noise
intensity at the steepest point of the response curve
(the inflection point $\sigma_c$, defined as
$\mathrm{argmax}_{\sigma}\,\mathrm{d}\langle H\rangle/\mathrm{d}\sigma$)
decreases with $N$.
For $N \in \{5, 8, 10, 15\}$ a preliminary power-law fit
to the inflection points gives
\begin{equation}
    \sigma_c(N) \approx C\,N^{-\alpha}, \qquad
    \alpha \approx 1.4 \quad (N \leq 15\text{ only}),
    \label{eq:powerlaw}
\end{equation}
as detailed above (four data points, $N \in \{5, 8, 10, 15\}$;
$R^2 = 0.96$ on the log-log fit).
Equation~\eqref{eq:powerlaw} is offered as an empirical
description of the $N \leq 15$ data only; we do not
extrapolate it to larger $N$, where the system
self-stabilizes and $\sigma_c$ loses physical meaning
(see below).
This exponent is obtained from the
\emph{inflection-point} definition of $\sigma_c$
(the noise intensity at steepest response).
An alternative \emph{threshold-crossing} definition
($\langle H \rangle_\mathrm{ss} > H_\mathrm{thr}$),
used in the universality test of
Section~\ref{subsec:universality} for computational
efficiency, yields $\alpha \approx 1.1$ at $\gamma = 1$;
the two values ($1.4$ vs $1.1$) are of the same order,
and the key qualitative result---that $\alpha$ depends on
the interaction form---is robust to the definition choice
(Section~\ref{subsec:universality}).
This approximate scaling breaks down at larger $N$.
At large $N$, the instability rate of the uniform
fixed point $\lambda = \gamma N^{-\gamma}$
(equation~\eqref{eq:eigenvalue}) vanishes:
for $\gamma = 1$, $\lambda = 1/N \to 0$, so the
deterministic ordering tendency weakens and the
system self-stabilizes without noise.
The analytical bound in~\ref{app:bound} confirms
this trend.


\subsection{Universality test: model-specificity of the scaling exponent}
\label{subsec:universality}

The power-law~\eqref{eq:powerlaw} raises a natural question:
does the exponent $\alpha \approx 1.4$ reflect a universal
property of diversity stabilization on the simplex,
or is it specific to the baseline interaction~\eqref{eq:baseline}?
To answer this, we replace the baseline drift with the
superlinear form~\eqref{eq:generalized} at $\gamma > 1$
and repeat the $\sigma_c(N)$ measurement for the same
set of system sizes, averaging over five independent trials
per parameter point to suppress single-run noise.

\begin{figure}[htbp]
    \centering
    \includegraphics[width=0.75\linewidth]{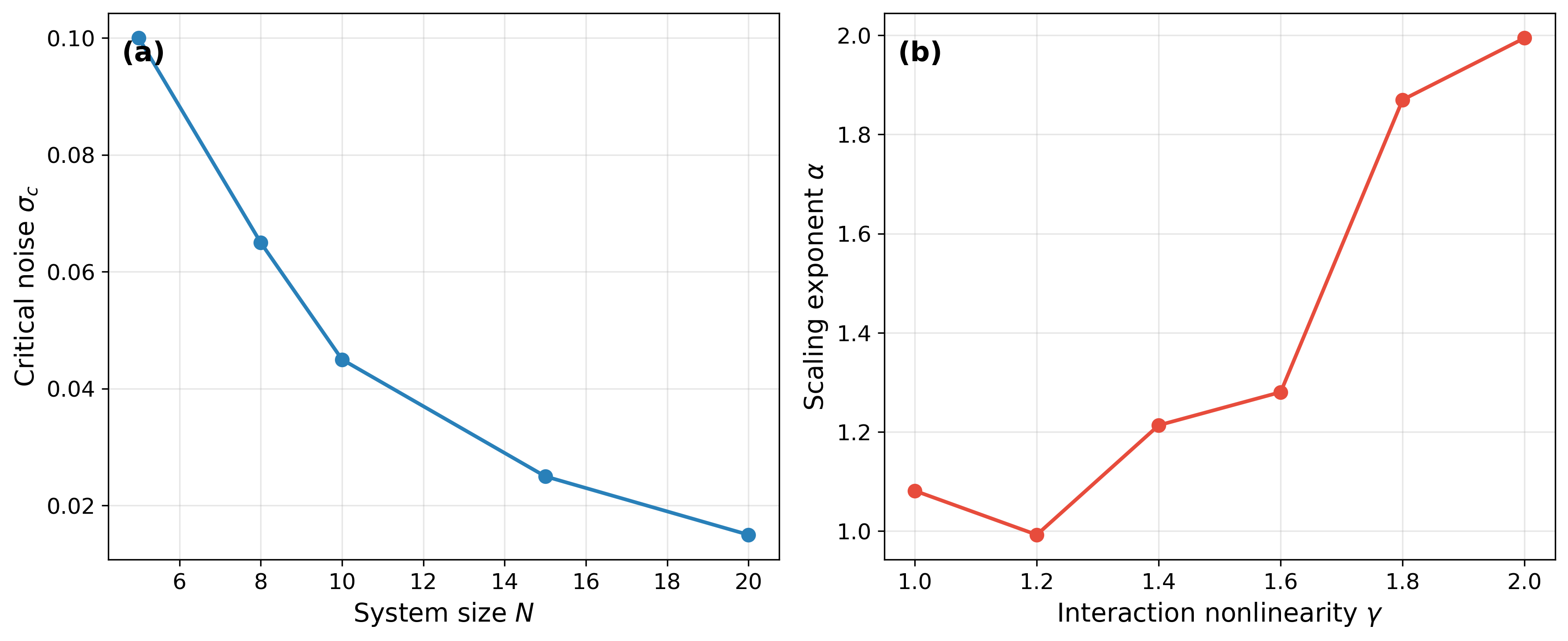}
    \caption{Model-specificity of the critical noise scaling.
    \textbf{(a)} Critical noise intensity $\sigma_c$ versus system
    size $N$ for the superlinear interaction~\eqref{eq:generalized}
    at $\gamma = 2$ (five trials per point).
    The curve is monotonically decreasing with fitted exponent
    $\alpha \approx 2.0$, steeper than the baseline value
    $\alpha \approx 1.1$.
    \textbf{(b)} Fitted scaling exponent $\alpha$ as a function of
    the nonlinearity parameter $\gamma$.
    The exponent remains near $\approx 1$ for $\gamma \leq 1.6$
    (with a non-monotonic dip at $\gamma = 1.2$), then rises
    to $\approx 2.0$ at $\gamma = 2.0$;
    $\alpha$ is model-dependent and sensitive to the
    interaction nonlinearity.}
    \label{fig:universality}
\end{figure}

Figure~\ref{fig:universality}a shows $\sigma_c(N)$ for
the superlinear interaction at $\gamma = 2$.
In contrast to the baseline model, the decay is steeper:
$\sigma_c$ drops from $\approx 0.10$ at $N = 5$ to
$\approx 0.015$ at $N = 20$, yielding a fitted exponent
$\alpha \approx 2.0$, steeper than the baseline value.
The curve is monotonically decreasing and well described
by a power law throughout.
The key observation is not a breakdown of the power-law
structure itself, but a systematic shift in the exponent:
stronger nonlinear coupling steepens the $\sigma_c(N)$
decay, meaning that larger systems become self-stabilizing
more rapidly under superlinear interactions.

To characterize how the exponent evolves as the nonlinearity
is increased continuously from the baseline, we fit
$\sigma_c(N)$ at each value of $\gamma \in
\{1.0, 1.2, 1.4, 1.6, 1.8, 2.0\}$ and plot the
resulting $\alpha(\gamma)$ in Figure~\ref{fig:universality}b.
The exponent remains near $\alpha \approx 1$ for
$\gamma \leq 1.6$ (ranging from $1.08$ at $\gamma = 1.0$
to $1.28$ at $\gamma = 1.6$, with a non-monotonic dip
to $0.99$ at $\gamma = 1.2$), then rises sharply to
$\alpha \approx 2.0$ at $\gamma = 2.0$.
The overall trend is an increase from $\approx 1.1$
to $\approx 2.0$, but the low-$\gamma$ plateau and
the non-monotonicity at $\gamma = 1.2$ indicate that
the $\gamma$-dependence of $\alpha$ is not a simple
smooth function.

These results indicate that $\alpha$ is a
\emph{model-dependent} quantity that varies
with the functional form of the competitive interaction,
rather than a universal critical index in the sense of
renormalization-group universality classes \cite{ref1,ref2}.
The sensitivity of $\alpha$ to $\gamma$ indicates that the
finite-size scaling of $\sigma_c$ reflects the specific way
the nonlinear drift balances positive feedback against the
simplex constraint at each interaction strength.
The structured $\gamma$-dependence of $\alpha$ is
partially captured by the analytical bound derived
in~\ref{app:bound}: linearization of the drift at
the uniform fixed point yields the prediction
$\alpha_\mathrm{theory}(\gamma) = (\gamma+2)/2$
(equation~\eqref{eq:alpha-theory}), which increases
linearly from $1.5$ at $\gamma = 1$ to $2.0$ at
$\gamma = 2$.
The numerical values lie systematically below this
prediction at low $\gamma$ (where the instability is
strong and linearization is least accurate) but
converge to it at high $\gamma$
($\alpha \approx 2.0$ vs $\alpha_\mathrm{theory} = 2.0$
at $\gamma = 2$), consistent with the weakening of the
instability rate $\lambda = \gamma\,N^{-\gamma}$ that
makes the linearized approximation progressively
more accurate.

\section{Nonlinear phase structure and absence of collective criticality}
\label{sec:results-phase}

\subsection{Dynamical phase diagram}
\label{subsec:phase-diagram}

We now turn to the broader parameter space spanned by the
nonlinearity exponent $\gamma$ and the noise strength $\sigma$.
For each point on a $14 \times 35$ grid covering
$\gamma \in [0.5, 2.0]$ and $\sigma \in [0, 0.08]$,
we compute the steady-state mean normalized entropy
$\langle H \rangle_\mathrm{ss}$ at $N = 20$, averaging
over five independent trials and the post-burn-in trajectory.
The result is displayed as a heat map in
Figure~\ref{fig:fluctuations}a.

The heat map reveals a systematic dependence of the
stationary diversity on both control parameters.
At low nonlinearity ($\gamma \lesssim 0.8$) the entropy
is generally depressed across the entire range of $\sigma$
explored, with the most pronounced ordering occurring at
small $\sigma$ where the deterministic collapse tendency
is least counteracted.
As $\gamma$ increases beyond approximately $1.0$--$1.4$,
the entropy rises progressively and eventually saturates
near its maximum value for $\gamma \gtrsim 1.5$,
largely irrespective of $\sigma$.
In this high-$\gamma$ regime the system maintains a
high-diversity stationary state even in the absence of
noise ($\sigma \approx 0$), indicating that the
superlinear interaction itself acts as a self-stabilizing
mechanism that substitutes for external stochastic forcing.

The transition between the low-entropy and
high-entropy regimes is continuous and gradual throughout
the parameter plane.
No sharp boundary, discontinuous jump, or narrow
coexistence region is visible in Figure~\ref{fig:fluctuations}a;
the equal-entropy contours run as smooth, broadly spaced
curves with no sign of bunching or abrupt curvature change.
Automated detection of the maximum entropy gradient over
the full grid yields a value of $0.041$ per unit $\gamma$,
confirming the absence of any sharp structural feature.
The grid spacing in $\gamma$ is approximately
$0.115$, so reported peak positions carry an intrinsic
resolution uncertainty of $\pm 0.06$; this does not
affect any qualitative conclusion but should be borne
in mind when comparing peak locations across experiments.
More generally, the precise location and amplitude of
the $\chi_H$ peak are sensitive to the random-seed
ensemble and the parameter grid, and vary by $\sim 0.2$
in $\gamma$ and $\sim 20\%$ in amplitude across
independent runs.
The conclusions of this section depend only on two
robust features that are stable across all runs:
(i)~$\chi_H$ does not diverge with $N$, and
(ii)~$\chi_H$ is strongly suppressed for
$\gamma \gtrsim 1.5$.
The system therefore exhibits a smooth \emph{crossover}
rather than a phase transition as $\gamma$ is varied,
a distinction with important consequences for the
fluctuation analysis presented in
Section~\ref{subsec:fluctuations}.


\subsection{Fluctuation susceptibility analysis}
\label{subsec:fluctuations}

\begin{figure*}[htbp]
    \centering
    \includegraphics[width=0.85\linewidth]{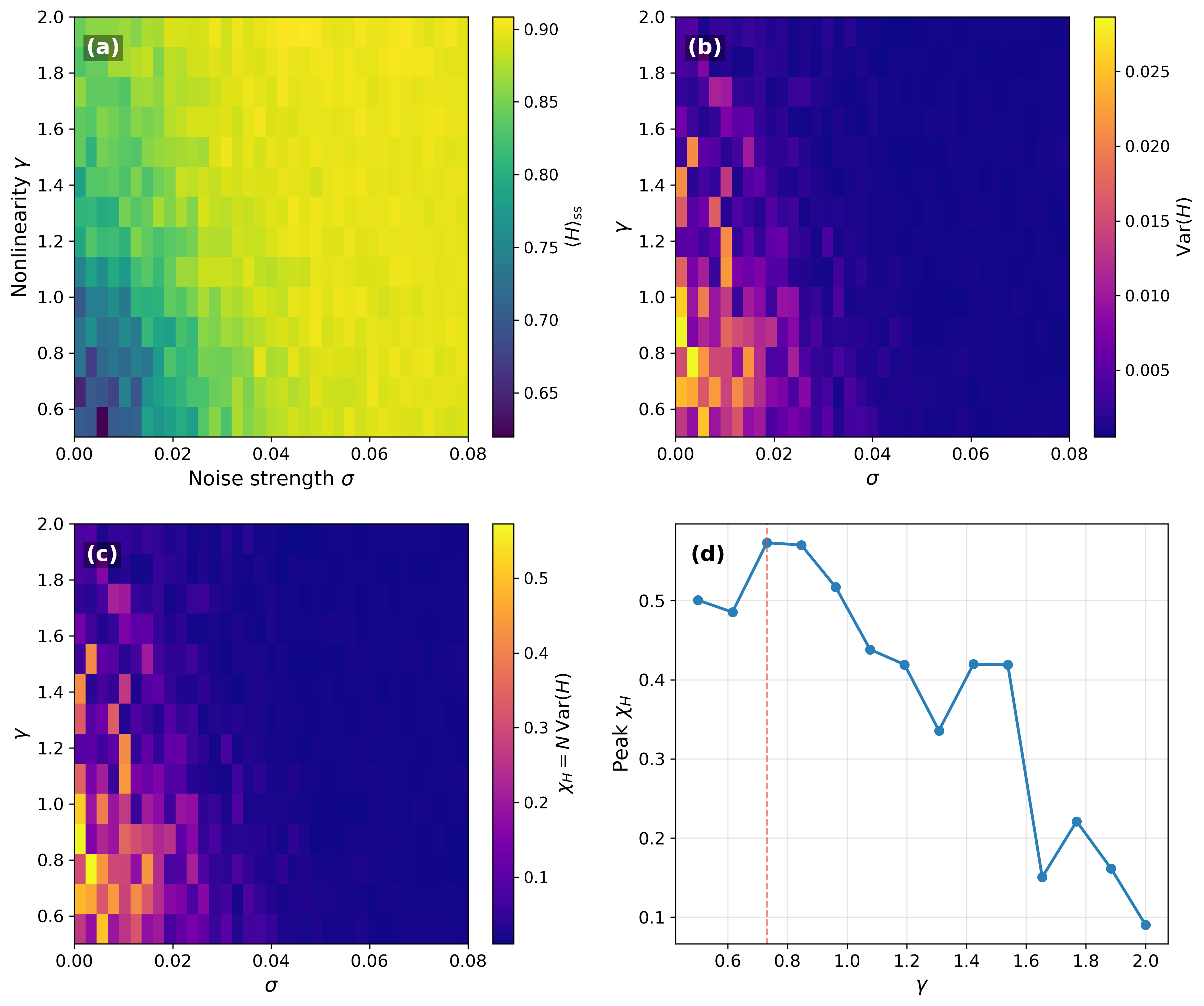}
    \caption{Phase structure and fluctuation analysis at $N = 20$.
    \textbf{(a)} Dynamical phase diagram in the
    $(\gamma, \sigma)$ plane; color encodes
    $\langle H \rangle_\mathrm{ss}$.
    The transition from low to high entropy is a smooth crossover.
    \textbf{(b)} Heat map of the entropy variance
    $\mathrm{Var}(H)$ in the $(\gamma, \sigma)$ plane.
    The global maximum occurs at low $\gamma \approx 0.7$;
    variance decreases as $\gamma$ increases.
    \textbf{(c)} Fluctuation susceptibility
    $\chi_H = N\,\mathrm{Var}(H)$ map; peak value
    $\chi_H \approx 0.6$
    at $(\gamma,\sigma)\approx(0.7, 0.002)$.
    \textbf{(d)} Peak susceptibility $\chi_H^\mathrm{peak}$
    versus nonlinearity $\gamma$ (maximum over all $\sigma$
    at each $\gamma$; trials$=20$).
    The curve falls on average from $\approx 0.6$ at
    $\gamma \approx 0.7$ to $\approx 0.10$ at $\gamma = 2.0$;
    no divergence or pronounced peak is observed near
    $\gamma \approx 1.4$, ruling out a critical phase
    transition at that location.
}
    \label{fig:fluctuations}
    \label{fig:phasediagram}
\end{figure*}

The smooth crossover seen in Figure~\ref{fig:fluctuations}a
raises the question of whether the structural change near
$\gamma \approx 1.0$--$1.4$ nevertheless harbors genuine
critical behavior at the level of fluctuations.
In equilibrium statistical mechanics, a second-order phase
transition that appears continuous in the order parameter
can still produce a divergent susceptibility at the
critical point \cite{ref2,ref4}.
We therefore compute the full fluctuation landscape
$\chi_H(\gamma, \sigma) = N\,\mathrm{Var}(H)$
over the same $14 \times 35$ parameter grid used in
Section~\ref{subsec:phase-diagram}.

\paragraph{Entropy variance and susceptibility maps.}
Figure~\ref{fig:fluctuations}b shows the variance
$\mathrm{Var}(H)$ as a heat map in the
$(\gamma, \sigma)$ plane.
The global maximum of $\mathrm{Var}(H)$ is located
in the low-$\gamma$ region near
$\gamma \approx 0.7$, $\sigma \approx 0.03$,
and the variance decreases progressively as $\gamma$
increases toward 2.0.
The susceptibility map $\chi_H = N\,\mathrm{Var}(H)$
shown in Figure~\ref{fig:fluctuations}c reproduces
the same spatial structure---as expected, since the
two quantities differ only by a factor of $N = 20$---with
a peak value of $\chi_H \approx 0.6$ at
$(\gamma, \sigma) \approx (0.7, 0.002)$.
Notably, neither map shows any concentrated bright region
near $\gamma \approx 1.4$, the location where the
phase diagram changes appearance most visibly.

\paragraph{Peak susceptibility as a function of $\gamma$.}
To make this point precise, Figure~\ref{fig:fluctuations}d
shows $\chi_H^\mathrm{peak}(\gamma)$, defined as the
maximum of $\chi_H$ over all $\sigma$ values at each
fixed $\gamma$.
The curve attains its global maximum of $\approx 0.6$ at
$\gamma \approx 0.7$ and decreases on average as
$\gamma$ increases, falling to $\approx 0.10$ by $\gamma = 2.0$
(a reduction by a factor of approximately six).
Local revivals appear at several intermediate $\gamma$ values,
but all remain well below the primary peak and are associated
with the stochastic variability of the low-$\gamma$ regime
identified in Section~\ref{subsec:robustness}.
There is no peak, kink, or inflection point of any
significance at $\gamma \approx 1.4$.

\paragraph{Absence of critical divergence: a key negative result.}
The foregoing observations constitute a direct test of
the criticality hypothesis.
If the crossover near $\gamma \approx 1.4$ were a genuine
second-order phase transition, the susceptibility
$\chi_H$ should exhibit a pronounced maximum or
divergence at the transition point, growing with
system size as $\chi_H \sim N^{\eta/\nu}$
in the spirit of finite-size scaling \cite{ref2,ref4}.
The data show the opposite: the susceptibility peaks at
$\gamma \approx 0.7$ with $\chi_H \approx 0.6$
and falls by a factor of approximately six by
$\gamma = 2.0$; the value at $\gamma \approx 1.4$
is less than one-third of the low-$\gamma$ maximum.
While the susceptibility peak at $\gamma \approx 0.7$
may superficially resemble a critical point, its lack of
divergence with $N$ (Section~\ref{subsec:robustness})
and the insensitivity of the relaxation timescale to $\gamma$
(Section~\ref{sec:results-relaxation}) are inconsistent
with equilibrium or nonequilibrium universality classes
\cite{ref1,ref2}.
We conclude that no critical phase transition occurs along
the $\gamma$ axis within the parameter range explored.
Instead, the system enters an
\emph{entropy-saturated nonlinear phase} for
$\gamma \gtrsim 1.5$, characterized by high mean entropy
and simultaneously low collective fluctuations---a
combination inconsistent with criticality but consistent
with a nonlinear smoothing mechanism, as elaborated in
Section~\ref{subsec:three-regimes}.

\subsection{Finite-size verification}
\label{subsec:robustness}

A finite-size check of the fluctuation suppression result
is provided in Supplementary Material~S6.
Using matched trial counts (trials$\,=20$ for all $N$),
the peak susceptibilities at $N \in \{10, 20, 40\}$
are $0.7$, $0.6$, and $0.5$ respectively
(see Supplementary Material~S6).
The values remain within the same order of magnitude
across all three sizes, with no systematic divergence
as $N$ increases.
This is inconsistent with the divergent growth
$\chi_H^\mathrm{peak} \sim N^{\eta/\nu}$ expected near
a genuine second-order phase transition \cite{ref4,ref5}.
Furthermore, in all three cases the high-$\gamma$ region
($\gamma \gtrsim 1.5$) shows strongly suppressed $\chi_H$,
confirming the fluctuation suppression identified in
Section~\ref{subsec:fluctuations} regardless of system size.

\subsection{Three-regime dynamical picture}
\label{subsec:three-regimes}

The results of Sections~\ref{subsec:phase-diagram}--\ref{subsec:robustness}
define three regimes as a function of $\gamma$
(see also Table~\ref{tab:regimes}).
\textbf{Regime~I} ($\gamma \lesssim 0.8$): the deterministic
drift biases the system toward low-entropy states;
$\chi_H$ is large ($\approx 0.6$) but does not diverge
with $N$, identifying a finite-size effect rather than
true criticality.
\textbf{Regime~II} ($0.8 \lesssim \gamma \lesssim 1.4$):
a smooth crossover in which the stationary probability mass
redistributes from the ordering corners toward the simplex
interior; $\chi_H$ declines without any local extremum.
\textbf{Regime~III} ($\gamma \gtrsim 1.5$): the mean entropy
saturates near unity independently of $\sigma$, and
$\chi_H$ is strongly suppressed---a hallmark of
fluctuation suppression by the nonlinear deterministic
coupling rather than by noise \cite{ref10}.
The dynamical consequences of this suppression are
analyzed in Section~\ref{sec:results-relaxation}.

\section{Relaxation dynamics and pathway multiplicity}
\label{sec:results-relaxation}

\subsection{Insensitivity of the relaxation timescale to nonlinearity}
\label{subsec:relaxation-structure}

The fluctuation analysis of Section~\ref{sec:results-phase}
characterizes the stationary state but does not reveal how
the system reaches it.
We now measure the entropy relaxation time
$\tau_\mathrm{relax}$ defined in equation~\eqref{eq:tau}
as a function of $\gamma$ at fixed $N = 20$
and $\sigma = 0.03$, a noise level within the range
where the susceptibility peak is observed.

\begin{figure}[htbp]
    \centering
    \includegraphics[width=0.75\linewidth]{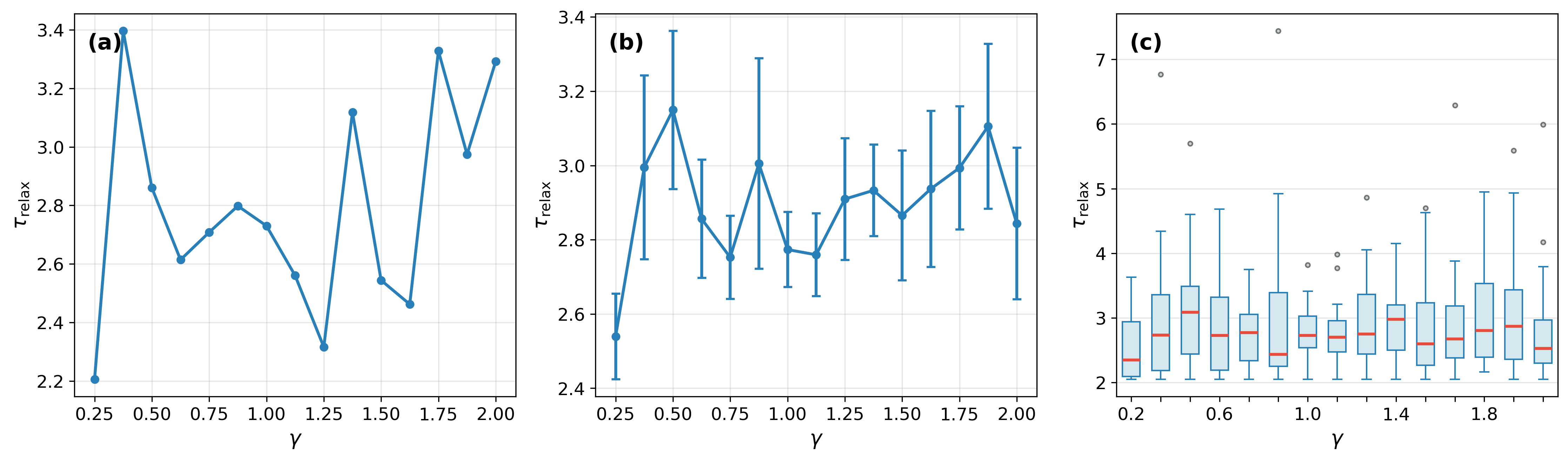}
    \caption{Entropy relaxation time $\tau_\mathrm{relax}$
    as a function of nonlinearity $\gamma$
    at $N = 20$, $\sigma = 0.03$.
    \textbf{(a)} Mean of five independent trials;
    the curve fluctuates between $2.2$ and $3.4$ time
    units with no systematic dependence on $\gamma$.
    \textbf{(b)} Mean $\pm$ one standard error over
    twenty independent trials.
    The grand mean is $\approx 2.9$ across all $\gamma$;
    the curve shows no monotonic trend, in contrast to
    the sixfold variation of $\chi_H$ over the same
    $\gamma$ range (Figure~\ref{fig:fluctuations}d).
    \textbf{(c)} Boxplot of per-trial relaxation times.
    The interquartile range is comparable across all
    $\gamma$ values, confirming that neither the central
    tendency nor the variability of $\tau_\mathrm{relax}$
    depends on nonlinearity strength.}
    \label{fig:relaxation}
\end{figure}

Figure~\ref{fig:relaxation}a shows $\tau_\mathrm{relax}(\gamma)$
computed from five independent trials.
The curve fluctuates between $2.2$ and $3.4$ time units
with no systematic trend as $\gamma$ increases from
$0.25$ to $2.0$.

To obtain statistically reliable estimates,
Figure~\ref{fig:relaxation}b shows $\tau_\mathrm{relax}$
averaged over twenty independent trials, with error bars
representing one standard error.
The mean curve confirms the absence of a systematic trend:
$\langle\tau_\mathrm{relax}\rangle$ fluctuates between
$2.5$ and $3.2$ with a grand mean of approximately $2.9$
across all $\gamma$ values, and no monotonic increase
or decrease is discernible.
The trial-to-trial spread is substantial throughout
(standard deviation $\approx 0.5$--$1.3$), reflecting
persistent variability at all nonlinearity strengths.

Figure~\ref{fig:relaxation}c displays the raw
per-trial relaxation times as a boxplot.
At every $\gamma$ the interquartile range spans
approximately $[2.0,\ 3.5]$, with occasional outliers
reaching $5$--$7$ time units.
The boxplot widths are comparable across the full
$\gamma$ range, confirming that the variability structure
is insensitive to nonlinearity strength.

The central finding of this section is therefore
\emph{negative}: $\tau_\mathrm{relax}$ does not
mirror the dramatic $\gamma$-dependence of $\chi_H$.
While the susceptibility decreases approximately
sixfold from low to high $\gamma$
(Section~\ref{subsec:fluctuations}),
the relaxation timescale remains essentially constant.

\subsection{Decoupling between relaxation rate and fluctuation suppression}
\label{subsec:mechanism}

The insensitivity of $\tau_\mathrm{relax}$ to $\gamma$,
combined with the strong $\gamma$-dependence of $\chi_H$,
constitutes the most informative result of the relaxation
analysis.
The decoupling between these two observables rules out
a simple potential-deepening mechanism for fluctuation
suppression and instead points to a pathway-level
explanation.

\paragraph{What the decoupling rules out.}
If the suppression of $\chi_H$ at high $\gamma$ arose
from a progressively deeper effective potential well
around the high-entropy stationary state---analogous to
increasing the spring constant in a harmonic trap---then
the relaxation time should decrease proportionally:
a deeper well implies faster convergence.
The data show the opposite: $\chi_H$ drops sixfold
while $\tau_\mathrm{relax}$ remains unchanged at
$\approx 2.9$ time units.
This rules out any mechanism in which the stationary
distribution narrows because the individual-trajectory
dynamics converge more rapidly to a single preferred
macrostate.

\paragraph{What the decoupling implies.}
The combination of constant $\tau_\mathrm{relax}$ and
decreasing $\chi_H$ has a natural interpretation in terms
of the law of total variance.
Conditioning on the microscopic relaxation pathway $\omega$,
\begin{equation}
    \mathrm{Var}(H) = \mathbb{E}\bigl[\mathrm{Var}(H\mid\omega)\bigr]
    + \mathrm{Var}\bigl(\mathbb{E}[H\mid\omega]\bigr).
    \label{eq:total-var}
\end{equation}
Because each trajectory takes approximately the same time
to reach stationarity regardless of $\gamma$, the rate
at which individual trajectories explore the configuration
space is not enhanced by nonlinearity.
What changes is the \emph{endpoint distribution}:
at high $\gamma$, different trajectories---following
qualitatively different microscopic paths (as evidenced
by the persistent trial-to-trial variability in
$\tau_\mathrm{relax}$)---converge to macroscopic states
with similar entropy values.
In the language of equation~\eqref{eq:total-var},
the second term
$\mathrm{Var}(\mathbb{E}[H\mid\omega])$ is suppressed
because different pathways $\omega$ produce similar
conditional means $\mathbb{E}[H\mid\omega]$, even
though the pathways themselves are diverse.

We term this
pathway-multiplicity-driven decorrelation:
strong nonlinear coupling does not accelerate the dynamics
or deepen a single basin of attraction, but rather creates
many approximately equivalent routes to the high-entropy
interior of the simplex.
The resulting diversity of pathways, combined with the
similarity of their endpoints, suppresses the ensemble
variance without affecting the trajectory-level
relaxation rate.

\paragraph{Falsifiability.}
The decoupling-based argument makes a specific,
testable prediction: fluctuation suppression via pathway
multiplicity requires $\tau_\mathrm{relax}$ to be
\emph{independent} of $\gamma$ while $\chi_H$ decreases.
If instead $\tau_\mathrm{relax}$ were found to decrease
proportionally with $\chi_H$ (i.e., both observables
tracking the same control parameter), this would indicate
a potential-deepening mechanism rather than pathway
multiplicity, and the present interpretation would be
falsified.
The data support the decoupling scenario.

\subsection{Mechanistic summary}
\label{subsec:mechanism-summary}

Table~\ref{tab:regimes} summarizes the joint behavior
of $\chi_H$ and $\tau_\mathrm{relax}$ across the three
dynamical regimes.

\begin{table}[htbp]
\centering
\caption{Summary of the three dynamical regimes.
$\chi_H^\mathrm{peak}$: peak fluctuation susceptibility
over $\sigma$ at each $\gamma$.
$\langle\tau\rangle$: mean relaxation time (20 trials).
$\mathrm{std}(\tau)$: trial-to-trial standard deviation.}
\label{tab:regimes}
\begin{tabular}{lccc}
\hline
\textbf{Regime ($\gamma$ range)}
  & \textbf{$\chi_H^\mathrm{peak}$}
  & \textbf{$\langle\tau\rangle$}
  & \textbf{std($\tau$)} \\
\hline
I: ordering-prone \; ($\lesssim 0.8$)
  & large ($\approx 0.6$) & $\sim 2.9$ & $\sim 0.8$ \\[3pt]
II: crossover \; ($0.8$--$1.4$)
  & declining & $\sim 2.9$ & $\sim 0.6$ \\[3pt]
III: entropy-saturated \; ($\gtrsim 1.5$)
  & small ($\lesssim 0.15$) & $\sim 2.9$ & $\sim 0.9$ \\
\hline
\end{tabular}

\smallskip
\begin{flushleft}
\textit{Mechanistic notes.}
\textbf{All regimes:} The relaxation timescale
$\langle\tau_\mathrm{relax}\rangle \approx 2.9$ is
approximately constant across all three regimes,
while $\chi_H$ varies sixfold.
This decoupling implies that fluctuation suppression in
Regime~III does not arise from faster convergence
(deeper potential well) but from pathway-multiplicity-driven
ensemble decorrelation: diverse microscopic trajectories
converge to similar macroscopic endpoints, reducing the
between-trajectory variance without affecting the
individual relaxation rate.
\end{flushleft}
\end{table}

The key feature of Table~\ref{tab:regimes} is the
contrast between the $\chi_H$ column, which varies
sixfold across regimes, and the $\langle\tau\rangle$
column, which remains constant.
This decoupling is the primary evidence for the
pathway-multiplicity mechanism: fluctuation suppression
in the high-$\gamma$ regime is not a consequence of
faster or more deterministic relaxation, but of
the convergence of diverse relaxation pathways to
similar macroscopic endpoints.
Together with the smooth crossover established in
Section~\ref{subsec:phase-diagram} and the absence
of critical fluctuations confirmed in
Section~\ref{subsec:fluctuations}, this defines
\emph{nonlinear fluctuation suppression via pathway
multiplicity} as the operative stabilization pattern
in this class of stochastic competitive systems.

\section{Discussion}
\label{sec:discussion}

\subsection{Synthesis of main findings}
\label{subsec:synthesis}

The small-$N$ scaling exponent $\alpha$
is a model-specific quantity tied to the baseline
replicator interaction~\eqref{eq:baseline}, not a
universal critical index \cite{ref1,ref2,ref3}:
replacing the interaction with a superlinear variant
shifts the exponent from $\approx 1.1$
to $\approx 2.0$, with a non-trivial $\gamma$-dependence.
The nonlinearity exponent $\gamma$ governs the
\emph{type} of collective behavior---ordering-prone
at low $\gamma$, smoothly crossing over to
entropy-saturated at high $\gamma$---without
producing a divergent susceptibility or a
$\gamma$-dependent relaxation timescale \cite{ref21,ref1,ref22}.

The core finding is that strong nonlinearity
suppresses collective fluctuations while leaving the
relaxation timescale unchanged---a decoupling that
rules out potential-deepening mechanisms and instead
points to pathway-multiplicity-driven ensemble
decorrelation as the operative stabilization pattern.
This pattern---\emph{nonlinear fluctuation suppression
via pathway multiplicity}---is distinct from
noise-induced stabilization \cite{ref10,ref18}
(which operates through stochastic forcing) and from
glassy dynamics \cite{ref13,ref15,ref16}
(where suppression accompanies slowing of relaxation).
This decoupling between
relaxation rate and fluctuation suppression has not been
previously reported in simplex-constrained competitive
dynamics;
existing accounts attribute diversity stabilization
either to noise \cite{ref10,ref22} or to deterministic
attractors \cite{ref11,ref14}, but not to an ensemble
decorrelation effect operating independently of both.
The regime structure identified here
(Table~\ref{tab:regimes}) is expected to arise in other
competitive-exclusion dynamics on the simplex, such as
Lotka--Volterra systems and cyclic competition models,
provided the interaction preserves the simplex constraint.
Beyond these model-level extensions, the regime structure
admits a qualitative correspondence with empirically
measurable quantities in real competitive systems,
offered here as an interpretive mapping rather than a
quantitative fit.
In microbial community ecology, for instance, the noise
intensity $\sigma$ plays the role of environmental
fluctuation strength and the normalized entropy $H$
corresponds directly to the Shannon diversity index
routinely measured in that field;
the low-$\gamma$, noise-dependent regime then maps onto
the empirically documented situation in which
environmental variability sustains species coexistence
that deterministic competition would eliminate, while
the high-$\gamma$, self-stabilizing regime maps onto
communities whose diversity is maintained by strong
niche structure largely independently of external noise.
An analogous reading applies to market-share dynamics,
with $1 - \mathrm{HHI}$ (one minus the
Herfindahl--Hirschman concentration index) as the
diversity observable.
These are qualitative correspondences intended to locate
the model's regimes within known phenomenology;
quantitative calibration against specific datasets is
left to future work.

\subsection{Relation to existing theoretical frameworks}
\label{subsec:theory-comparison}

\paragraph{Noise-induced transitions.}
The theory of noise-induced transitions developed by
Horsthemke and Lefever \cite{ref10} identifies conditions
under which additive or multiplicative noise shifts the
stationary distribution of a nonlinear system, sometimes
creating new phases that have no deterministic counterpart.
In that framework, the noise strength $\sigma$ plays the
role of a bifurcation parameter, and a transition occurs
when the stationary probability density changes its
qualitative shape---for instance, from unimodal to
bimodal---at a critical $\sigma_c$.
The present system shares the qualitative feature that
noise promotes a more uniform (high-entropy) stationary
distribution, and the critical noise intensity
$\sigma_c(N)$ identified in Section~\ref{sec:results-scaling}
is formally analogous to a noise-induced transition threshold.
However, the fluctuation analysis of
Section~\ref{sec:results-phase} reveals a key difference:
the ``transition'' is not accompanied by any divergence
of the susceptibility $\chi_H$, and its apparent location
in the $(\gamma, \sigma)$ plane shifts continuously with
system size rather than converging to a well-defined
thermodynamic limit.
Similarly, Van den Broeck and colleagues showed that
noise can induce genuine nonequilibrium phase transitions
with associated critical behavior in spatially extended
systems \cite{ref22}; the present model, being a
mean-field system on the simplex with no spatial
structure, does not develop the long-range correlations
necessary to sustain such a transition.
The diversity stabilization observed here is therefore
better described as a noise-biased smooth crossover \cite{ref23}
than as a noise-induced phase transition.

\paragraph{Finite-size scaling theory.}
Standard finite-size scaling \cite{ref4,ref5,ref6}
provides a systematic framework for extrapolating
finite-system measurements to the thermodynamic limit
near a genuine second-order transition.
At a critical point, the susceptibility scales as
$\chi \sim N^{\eta/\nu}$, and this divergence
can be detected even in small systems by tracking
how the peak susceptibility grows with $N$.
The results of Section~\ref{subsec:robustness}
demonstrate that this diagnostic fails in the present
system: matched-trial ($\mathrm{trials}=20$) measurements
at $N \in \{10, 20, 40\}$ yield peak values of
$0.7$, $0.6$, and $0.5$---values within the same
order of magnitude, with a slight \emph{decrease} with $N$.
This failure of FSS is itself informative---it
identifies the fluctuation peak at low $\gamma$ as
a finite-size pseudo-critical artefact, present because
small systems have relatively large boundary-to-bulk
ratios and are easily destabilized by fluctuations,
but diluted as the interior bulk behavior comes to
dominate at larger $N$.
The implication for numerical studies of similar
systems is methodological: a susceptibility peak
observed at a single system size is not sufficient
evidence for a phase transition; monotonic growth
of the peak with $N$ is the necessary condition,
and this test should be applied routinely.

\paragraph{Evolutionary game dynamics and large-population limits.}
In finite-population evolutionary game theory,
stochastic effects arising from demographic noise
are known to alter the long-run frequency distribution
of strategies relative to the deterministic replicator
prediction \cite{ref11,ref14}.
Taylor and Nowak \cite{ref14} showed that in finite
populations the probability of fixation---the analog
of diversity collapse in the present model---depends
sensitively on population size $N$, and that for
certain game structures larger populations are
effectively more deterministic, converging toward
the replicator equilibrium.
The present model exhibits a superficially similar
large-$N$ trend---$\sigma_c \to 0$ as $N \to \infty$---but
the mechanism differs.
In evolutionary game theory the noise is intrinsic
(demographic), scales as $N^{-1/2}$, and its effect
diminishes as $N$ grows because the law of large
numbers suppresses relative fluctuations.
Here the noise is external and additive, and the
diminishing requirement for noise as $N$ increases
reflects a growing self-organizing capacity of the
deterministic simplex dynamics, not a dilution of
stochastic effects.
The power-law exponent $\alpha$ quantifies
this self-organizing efficiency for the baseline
replicator interaction, and its sensitivity to
the interaction form (Section~\ref{subsec:universality})
suggests that the large-$N$ self-stabilization is
a model-specific property rather than a generic
feature of competitive dynamics on the simplex.

\subsection{Limitations and outlook}
\label{subsec:limitations}

The scaling exponent $\alpha$ applies only for
$N \leq 15$; at larger $N$ the system self-stabilizes.
The analytical prediction
$\alpha_\mathrm{theory} = (\gamma+2)/2$
(\ref{app:bound}) captures the trend; a rigorous
bound awaits nonlinear analysis of the boundary effects.

The two main conclusions---absence of a phase transition
and decoupling between $\tau_\mathrm{relax}$ and
$\chi_H$---are established numerically.
Analytical derivations (e.g.\ via Fokker--Planck
analysis of the stationary distribution) would
strengthen them and are a priority for future work.
The decoupling is an empirical observation consistent
with pathway multiplicity; a constructive proof that
pathway multiplicity implies decoupling remains open,
and the falsifiability criterion in
Section~\ref{subsec:mechanism} provides a testable
prediction for future studies.

Natural extensions include:
(i) Fokker--Planck analysis \cite{ref9} of the
stationary density $p_\mathrm{ss}(\mathbf{x})$
to obtain a sharper $\alpha$ prediction;
(ii) large-deviation theory \cite{ref19} to
characterize diversity collapse as a rare event;
(iii) graph-structured interactions to test whether
pathway multiplicity survives under spatial
heterogeneity;
and (iv) systematic variation of interaction classes
(asymmetric matrices, frequency-dependent selection)
to delineate the basin for which $\sigma_c(N)$
scaling is well-defined.

\section{Conclusion}
\label{sec:conclusion}

We have studied stochastic diversity stabilization in a
class of replicator-like competitive dynamics on the
$N$-dimensional probability simplex, parameterized by
an additive noise strength $\sigma$ and a nonlinearity
exponent $\gamma$.
Under the baseline linear replicator interaction,
the critical noise intensity required to sustain a
diverse stationary state decays as a power law,
$\sigma_c \sim N^{-\alpha}$, implying that sufficiently
large systems self-stabilize without external noise.
Replacing the baseline with a superlinear variant
shifts the exponent from $\alpha \approx 1.1$ to
$\approx 2.0$ as $\gamma$ increases, in quantitative
agreement with the linear-stability prediction
$\alpha = (\gamma+2)/2$ and establishing that $\alpha$
is a model-dependent quantity rather than a universal
critical index.
Scanning the full $( \gamma, \sigma )$ parameter plane
reveals a smooth dynamical crossover from an
ordering-prone, noise-dependent regime at low $\gamma$
to an entropy-saturated, self-organizing phase at
high $\gamma$, with no divergent fluctuation
susceptibility at any point along the transition:
the peak of $\chi_H = N\,\mathrm{Var}(H)$ is located
at $\gamma \approx 0.7$, well separated from the
visual crossover region near $\gamma \approx 1.4$,
and matched finite-size analysis finds no systematic
divergence with $N$, ruling out a genuine second-order
phase transition in the thermodynamic limit.
Relaxation dynamics complete the picture: the entropy
relaxation time $\tau_\mathrm{relax}(\gamma)$ is
approximately constant at $\approx 2.9$ time units
across the full $\gamma$ range, even as the susceptibility
decreases sixfold.
This decoupling between relaxation rate and fluctuation
amplitude rules out a simple potential-deepening mechanism
and implies that the system reaches the high-entropy
stationary state via a broad ensemble of diverse
microscopic pathways whose endpoints converge at
high $\gamma$.
The joint behavior of $\chi_H$ and $\tau_\mathrm{relax}$
supports a unified mechanistic interpretation: strong
nonlinear coupling suppresses collective entropy
fluctuations not by accelerating convergence toward
a single attractor, but by generating pathway multiplicity
that decorrelates trajectory endpoints at the ensemble
level---a mechanism we term \emph{nonlinear fluctuation
suppression via pathway multiplicity}, which is distinct
from both noise-induced transitions and glassy dynamics.

\appendix

\section{Failure of the microscopic relaxation criterion}
\label{app:criterion}

As noted in Section~\ref{subsec:observables},
the entropy relaxation time $\tau_\mathrm{relax}$
must be defined via a macroscopic convergence criterion
rather than a microscopic one.
This appendix documents the failure of the microscopic
approach in order to justify the choice made in
equation~\eqref{eq:tau}.

A natural first attempt is to declare relaxation complete
when the pointwise change in the state vector falls
below a threshold,
\begin{equation}
    \frac{1}{N}\sum_{i=1}^{N}
    \bigl|x_i(t) - x_i(t - \Delta t)\bigr|
    < \varepsilon_\mathrm{micro},
    \label{eq:micro-criterion}
\end{equation}
with $\varepsilon_\mathrm{micro} = 10^{-4}$.
In a noise-free system this criterion correctly
identifies the approach to a fixed point.
In the presence of additive Wiener noise, however,
the increment $x_i(t) - x_i(t-\Delta t)$ contains
a stochastic contribution of order
$\sigma\sqrt{\Delta t} \sim 10^{-3}$--$10^{-2}$
for the parameter values used here, which permanently
exceeds $\varepsilon_\mathrm{micro} = 10^{-4}$
regardless of how close the macroscopic distribution
is to stationarity \cite{ref7,ref8}.
As a consequence, criterion~\eqref{eq:micro-criterion}
never triggers, and every simulation run terminates
at the maximum allowed time $T = 6000$ steps,
corresponding to simulation time $t = 60$.

Applying this criterion to the same parameter settings
used in Section~\ref{sec:results-relaxation}
($N = 20$, $\sigma = 0.03$, $\gamma \in [0.25, 2.0]$)
produces a perfectly flat $\tau_\mathrm{relax} = 60$
for all $\gamma$---the maximum allowed simulation time---confirming
that the microscopic criterion carries no physical information
in any noise-driven system.
The macroscopic entropy criterion~\eqref{eq:tau}
resolves this issue: the resulting
$\tau_\mathrm{relax} \approx 2.9$ is two orders of
magnitude below the ceiling, confirming genuine convergence
to stationarity (Figure~\ref{fig:relaxation}).


\section{Analytical bound on $\sigma_c(N,\gamma)$}
\label{app:bound}

We derive a scaling bound on the critical noise intensity
as a function of both system size $N$ and nonlinearity
exponent $\gamma$, using the linear stability
result~\eqref{eq:eigenvalue} from Section~\ref{subsec:model}.

\subsubsection*{Step 1: Heuristic stationary variance}

The uniform fixed point $x_i = 1/N$ is linearly unstable
with instability rate $\lambda = \gamma\,N^{-\gamma}$
(equation~\eqref{eq:eigenvalue}).
In a purely linear system, this instability would cause
unbounded growth; in the present model, the simplex
boundaries provide a reflecting constraint that confines
the dynamics and permits a stationary distribution.
As a heuristic estimate, we balance the outward drift
(rate $\lambda$), the diffusive spreading (intensity $\sigma$),
and the confining effect of the simplex boundaries
to obtain the characteristic displacement scale,
yielding a variance estimate
\begin{equation}
    \mathrm{Var}(x_i) \;\sim\;
    \frac{\sigma^2}{2\lambda}
    = \frac{\sigma^2\,N^{\gamma}}{2\gamma}.
    \label{eq:var-general}
\end{equation}
Equation~\eqref{eq:var-general} is an
order-of-magnitude estimate, not a rigorous result:
the uniform point is unstable, so the standard
Ornstein--Uhlenbeck stationary-variance formula does
not strictly apply, and the true stationary distribution
is shaped by the nonlinear drift and the simplex
boundaries.
The estimate is nevertheless useful because it correctly
captures the parametric dependence on $N$ and $\gamma$.
All conclusions drawn from this appendix rest solely on
the resulting scaling exponent
$\alpha_\mathrm{theory} = (\gamma+2)/2$ and not on the
prefactor: the simplex boundary, not the linearized rate
$\lambda$, sets the true confining scale, so the
$1/(2\lambda)$ factor should be read as fixing the
$N$- and $\gamma$-dependence rather than the absolute
coefficient.

\subsubsection*{Step 2: Entropy lower bound}

A second-order Taylor expansion of the normalized Shannon
entropy $H$~\eqref{eq:entropy} around
$\mathbb{E}[x_i] = 1/N$ gives, using
$|\partial^2 H / \partial x_i^2|_{1/N} = N/\ln N$,
\begin{equation}
    \mathbb{E}[H(\mathbf{x})] \;\gtrsim\; 1
    - \frac{N^2}{2\ln N}\,\mathrm{Var}(x_i)
    = 1 - \frac{\sigma^2\,N^{\gamma+2}}{4\gamma\ln N}.
    \label{eq:Hbound-general}
\end{equation}
Requiring $\mathbb{E}[H] \geq H_\mathrm{thr}$ and inverting:
\begin{equation}
    \sigma_c \;\lesssim\;
    2\sqrt{\frac{\gamma\,(1 - H_\mathrm{thr})\,\ln N}
    {N^{\gamma+2}}}
    \;\sim\; \sqrt{\gamma}\;N^{-(\gamma+2)/2}\,(\ln N)^{1/2}.
    \label{eq:bound-general}
\end{equation}
The effective scaling exponent is therefore
\begin{equation}
    \alpha_\mathrm{theory}(\gamma) = \frac{\gamma+2}{2},
    \label{eq:alpha-theory}
\end{equation}
which predicts a \emph{linear increase} of $\alpha$ with
$\gamma$.

\subsubsection*{Comparison with numerical results}

Equation~\eqref{eq:alpha-theory} makes two testable
predictions.
First, for the baseline interaction ($\gamma = 1$),
$\alpha_\mathrm{theory} = 3/2 = 1.5$;
the numerical estimates $\alpha \approx 1.4$
(inflection-point definition) and $\alpha \approx 1.1$
(threshold-crossing definition) are of the same order
of magnitude, with the discrepancy attributable to
finite-$N$ effects and the heuristic nature
of~\eqref{eq:var-general}.
Second, for $\gamma = 2$, $\alpha_\mathrm{theory} = 2.0$;
the threshold-crossing fit gives $\alpha \approx 2.0$,
in close quantitative agreement.
More generally, the predicted linear trend
$\alpha = (\gamma+2)/2$ captures the observed increase
of $\alpha$ with $\gamma$
(Figure~\ref{fig:universality}b), with the analytical
values systematically above the numerical ones at low
$\gamma$ and converging at high $\gamma$.
This convergence at high $\gamma$ is consistent with
the weakening of the instability rate
$\lambda = \gamma\,N^{-\gamma}$, which makes the
linearized approximation more accurate when the
uniform state is only marginally unstable.

\section*{Declaration of Competing Interest}

The author declares that they have no known competing financial
interests or personal relationships that could have appeared to
influence the work reported in this paper.

\section*{Data Availability Statement}

All simulation code (Python scripts for Figures~1--5 and
Supplementary Figures S4--S6), raw output data, and
analysis notebooks are publicly available at
\url{https://github.com/ouruocun-dotcom/stochastic-diversity-stabilization}.
The repository includes fixed random seeds, parameter files,
and instructions to reproduce every figure in this paper.

\section*{Acknowledgements}

The author would like to express sincere gratitude to the host
institution for the academic environment and support provided
during this research.
This work was conducted independently by the author.
The author used an AI language model (Claude, Anthropic) for
assistance with manuscript writing and language editing.
All scientific content, experimental design, data analysis,
and conclusions are the sole responsibility of the author.


\clearpage
\section*{Supplementary Material}
\setcounter{figure}{3}
\renewcommand{\thefigure}{S\arabic{figure}}
This document contains three supplementary figures referenced in the
main text.
Figure~\ref{fig:s4} provides the fluctuation-based dynamical boundary
extracted from the susceptibility landscape of the main text.
Figure~\ref{fig:s5} documents the numerical stability of the main
results with respect to the integration time step.
Figure~\ref{fig:s6} presents the matched-trial finite-size
verification of fluctuation suppression across
$N \in \{10, 20, 40\}$.

\section*{S4\quad Fluctuation-based dynamical boundary}

\begin{figure}[htbp]
    \centering
    \includegraphics[width=0.75\linewidth]{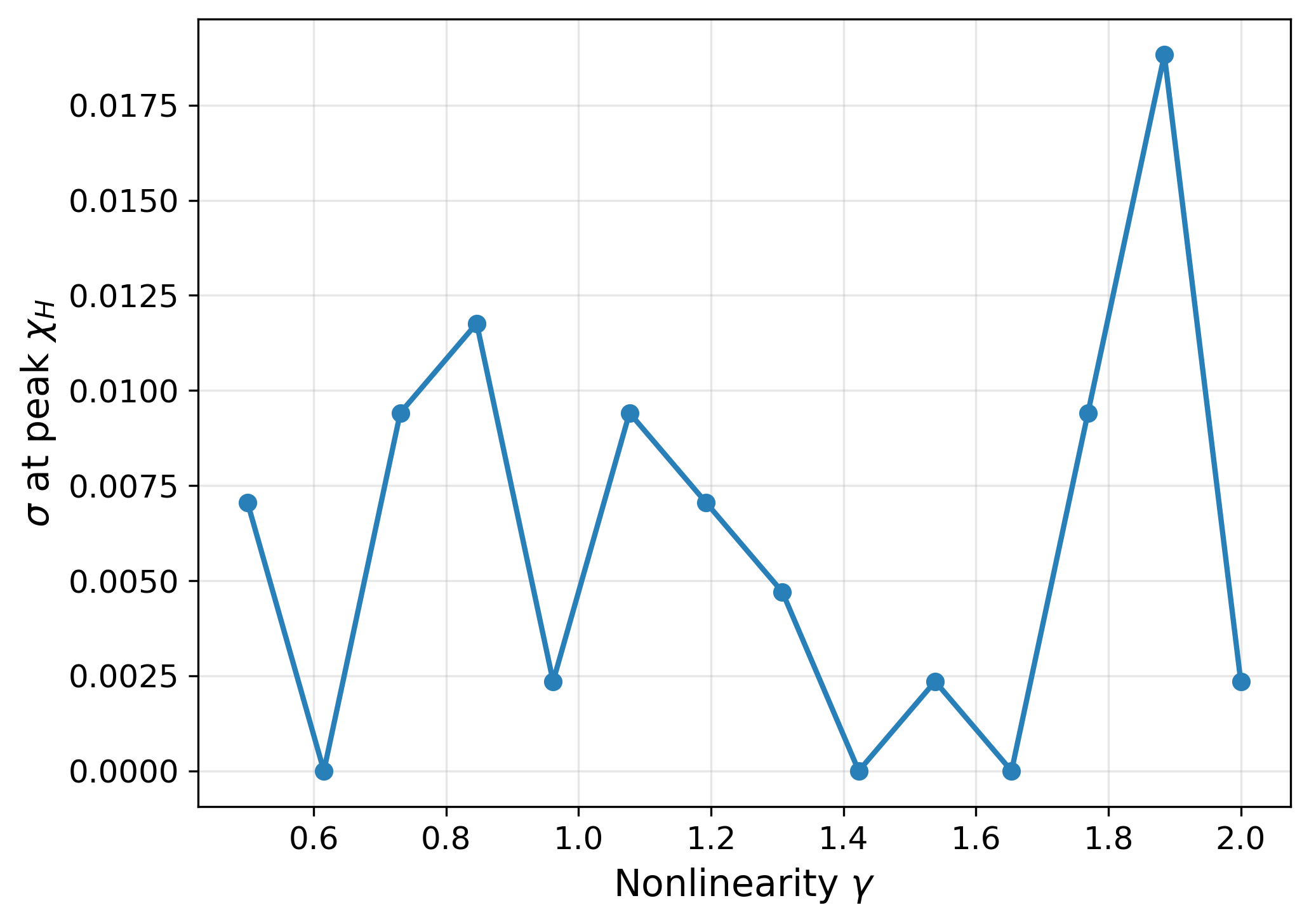}
    \caption{Fluctuation-based dynamical boundary $\sigma^*(\gamma)$,
    defined as the noise intensity at which $\chi_H$ is maximized
    over $\sigma$ at each fixed $\gamma$ ($N = 20$).
    The boundary is approximately stable for $\gamma \lesssim 1.2$
    but becomes highly erratic for $\gamma \gtrsim 1.3$,
    where the susceptibility landscape is nearly flat and
    the apparent maximum is dominated by statistical noise.
    This instability is a consequence of fluctuation suppression
    in the high-$\gamma$ regime (the fluctuation analysis of the main text)
    rather than a physical feature of the system.}
    \label{fig:s4}
\end{figure}

For each value of the nonlinearity exponent $\gamma$, the noise
intensity $\sigma^*(\gamma)$ at which the fluctuation susceptibility
$\chi_H = N\,\mathrm{Var}(H)$ attains its maximum over $\sigma$
defines a fluctuation-based dynamical boundary in the
$(\gamma, \sigma)$ plane.
Figure~\ref{fig:s4} shows this boundary at $N = 20$.

In the low-$\gamma$ regime ($\gamma \lesssim 1.2$), the boundary
is relatively stable, with $\sigma^*$ confined to a narrow
low-noise band and showing no clear trend.
For $\gamma \gtrsim 1.3$, the boundary becomes highly erratic,
with $\sigma^*$ swinging abruptly between near-zero values and
the upper edge of the scanned range without settling at any
stable value.
This instability does not reflect a physical phase boundary;
it arises because the susceptibility in the high-$\gamma$ regime
is uniformly low and nearly flat across all $\sigma$
(see Figure~4c of the main text), so the location of its maximum
is dominated by statistical noise rather than any structural feature.
The behavior is consistent with the fluctuation suppression
of the entropy-saturated regime identified in the main text,
and underscores that the fluctuation-based boundary is
not a meaningful diagnostic in the entropy-saturated phase.

\section*{S5\quad Numerical stability with respect to integration
time step}

\begin{figure}[htbp]
    \centering
    \includegraphics[width=\linewidth]{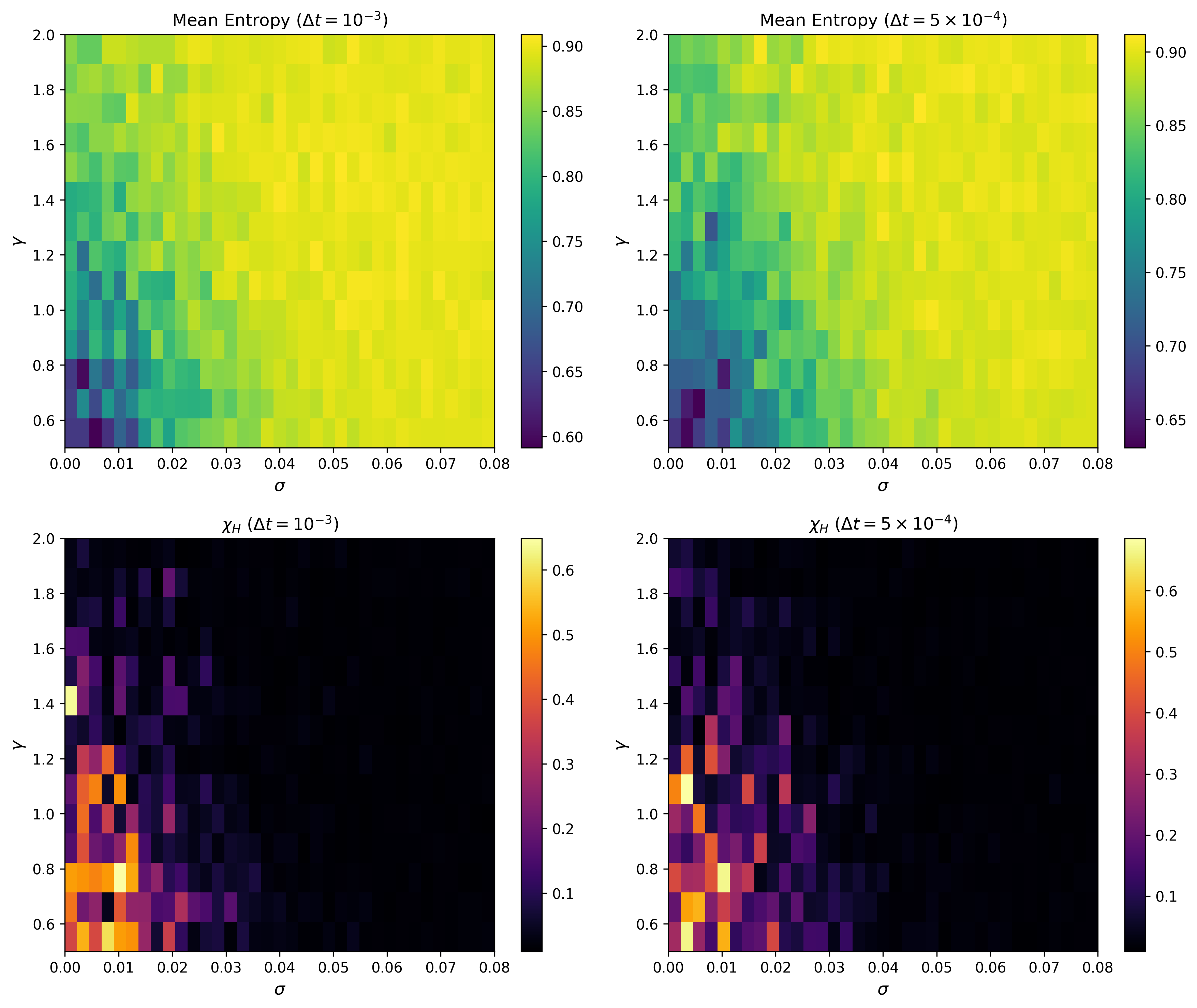}
    \caption{Numerical stability check: comparison of results at
    $\Delta t = 10^{-3}$ (left column) and
    $\Delta t = 5 \times 10^{-4}$ (right column), both at $N = 20$.
    \textbf{Top row:} steady-state mean normalized entropy
    $\langle H \rangle_\mathrm{ss}$ in the $(\gamma, \sigma)$ plane;
    color ranges differ by less than $2\%$.
    \textbf{Bottom row:} fluctuation susceptibility
    $\chi_H = N\,\mathrm{Var}(H)$; the qualitative structure---peak
    at low $\gamma$, suppression at high $\gamma$---is reproduced at
    both time steps.
    The shift in peak $\gamma$ position reflects
    seed- and grid-dependent variability documented
    in the main text, not numerical error.}
    \label{fig:s5}
\end{figure}

All main-text results use the integration time step
$\Delta t = 10^{-3}$.
To verify that the conclusions are not sensitive to this choice,
we repeat the phase diagram and fluctuation susceptibility
analysis at $\Delta t = 5 \times 10^{-4}$ and compare the
two outputs directly.
Figure~\ref{fig:s5} shows the mean entropy
$\langle H \rangle_\mathrm{ss}$ (top row) and the
susceptibility $\chi_H$ (bottom row) for both time steps
at $N = 20$.

The mean entropy maps (top row) are qualitatively identical:
both show the same progression from low entropy at small $\gamma$
to high, $\sigma$-independent entropy at large $\gamma$,
with the same approximate location of low-entropy pockets
near $(\gamma, \sigma) \approx (0.6, 0.00)$.
The color ranges differ by less than $2\%$ in the dynamic range.

The susceptibility maps (bottom row) confirm that the
spatial structure of $\chi_H$---peak at low $\gamma$,
suppression at high $\gamma$---is reproduced at both
time steps.
The peak location shifts between the two time steps,
consistent with the seed- and grid-dependent
peak-position variability documented in the main text,
rather than indicating integration error.
All qualitative conclusions of the main text are
unaffected by this variation.

\paragraph{Note on the relaxation time step.}
The relaxation analysis of the main text
uses a coarser step $\Delta t = 0.01$ to achieve a total
simulation time of 60 units within 6000 steps.
A separate $\Delta t$ convergence check is not required
for the relaxation observable, for two reasons.
First, $\tau_\mathrm{relax}$ is defined via a
running-window average of the macroscopic entropy
over 100-step windows (= 1.0 time unit at $\Delta t = 0.01$),
which smooths out step-level integration errors by
averaging over $\sim N \times 100$ individual entropy
contributions.
Second, the measured $\tau_\mathrm{relax} \approx 2.9$
time units is nearly three times the window size,
so the convergence criterion responds to the slow
macroscopic dynamics rather than to the step-level
numerical scheme.
The key requirement for the relaxation experiment is
sufficient total simulation time (60 units $\gg$
$\tau_\mathrm{relax} \approx 2.9$), not step-level
precision.
To confirm this directly, we repeated the relaxation
analysis at $\Delta t = 0.005$ (12000 steps, same total
time of 60 units, with the running window widened to
200 steps to preserve the 1.0 time-unit window width).
The grand mean over all $\gamma$ is
$\langle\tau_\mathrm{relax}\rangle = 2.89$ at
$\Delta t = 0.01$ and $2.81$ at $\Delta t = 0.005$,
a relative difference of only $3\%$.
Individual $\gamma$ points differ by up to
$\sim\!13\%$, but these differences are unsystematic---%
some points increase and others decrease between the
two time steps---and lie within the trial-to-trial
standard deviation reported in the relaxation analysis of the main text
($\mathrm{std}(\tau) \approx 0.5$--$1.3$).
The near-cancellation of these fluctuations in the
grand mean indicates that the systematic $\Delta t$
effect is negligible once trial-level scatter is
averaged out.
The qualitative conclusion---that $\tau_\mathrm{relax}$
is approximately constant across $\gamma$---holds
identically at both time steps.
This check is numerical (a grand-mean comparison) and
is not presented as a separate figure; the reproducing
script is provided in the code repository.

\section*{S6\quad Finite-size verification of fluctuation suppression}

\begin{figure}[htbp]
    \centering
    \includegraphics[width=\linewidth]{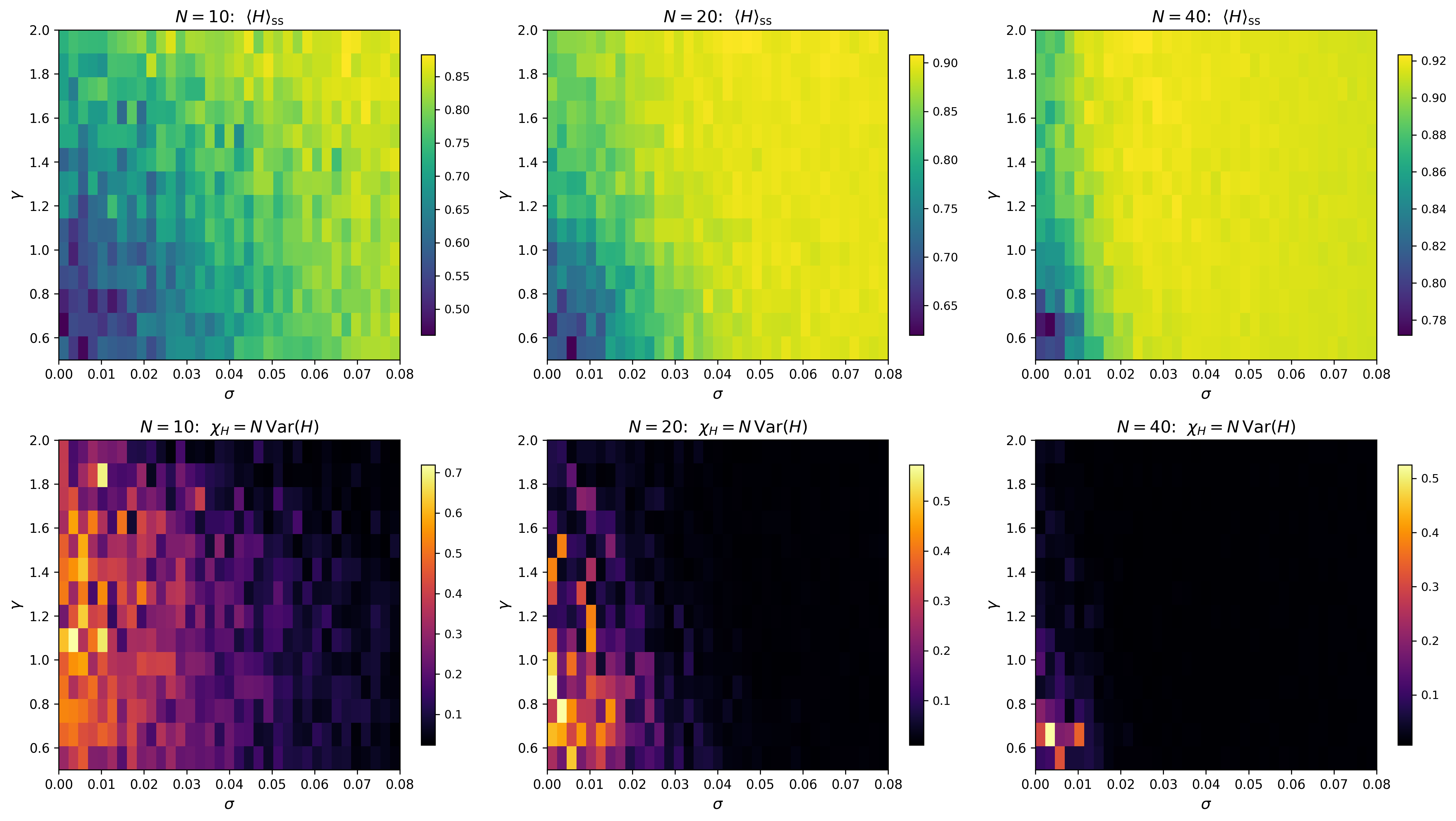}
    \caption{Finite-size verification of fluctuation suppression.
    Each column corresponds to a system size
    $N \in \{10, 20, 40\}$ (left to right).
    \textbf{Top row:} steady-state mean normalized entropy
    $\langle H \rangle_\mathrm{ss}$ in the $(\gamma, \sigma)$ plane.
    \textbf{Bottom row:} fluctuation susceptibility
    $\chi_H = N\,\mathrm{Var}(H)$.
    All three scans use trials$\,=20$.
    Peak values are $\approx 0.7$ ($N=10$, $\gamma \approx 1.1$),
    $\approx 0.6$ ($N=20$, $\gamma \approx 0.7$), and $\approx 0.5$ ($N=40$,
    $\gamma \approx 0.6$); no systematic increase with $N$ is
    observed, ruling out critical divergence \cite{ref4,ref5}.
    In all three panels, $\chi_H$ is strongly suppressed
    at $\gamma \gtrsim 1.5$, confirming size-independent
    fluctuation suppression in the high-nonlinearity regime.}
    \label{fig:s6}
\end{figure}

To assess whether the fluctuation suppression observed at $N = 20$
persists as the system size grows, we repeat the
$(\gamma, \sigma)$ susceptibility scan at $N \in \{10, 20, 40\}$.
Figure~\ref{fig:s6} shows the mean entropy
$\langle H \rangle_\mathrm{ss}$ (top row) and the fluctuation
susceptibility $\chi_H = N\,\mathrm{Var}(H)$ (bottom row)
for all three sizes, each computed with trials$\,=20$.

\paragraph{Results with matched trial counts.}
All three system sizes were scanned with trials$\,= 20$.
The peak susceptibilities are:
\begin{center}
\begin{tabular}{lcc}
\hline
$N$ & $\chi_H^\mathrm{peak}$ & peak $\gamma$ \\
\hline
10 & 0.7 & 1.08 \\
20 & 0.6 & 0.7 \\
40 & 0.5 & 0.62 \\
\hline
\end{tabular}
\end{center}
The three values lie within the same order of magnitude
and do not show a systematic increase with $N$;
indeed, the peak decreases slightly from $N=10$ to $N=40$.
A genuine second-order phase transition would require
$\chi_H^\mathrm{peak}$ to grow as $N^{\eta/\nu}$ \cite{ref4,ref5};
the absence of such growth rules out critical behavior.

\paragraph{High-$\gamma$ suppression is size-independent.}
In all three panels (Figure~\ref{fig:s6}, bottom row),
the susceptibility in the high-$\gamma$ regime
($\gamma \gtrsim 1.5$) is strongly suppressed relative
to the low-$\gamma$ peak, irrespective of system size.
This confirms that the fluctuation suppression identified
in the fluctuation analysis of the main text is a robust feature
of the high-nonlinearity regime and not a finite-size artefact.

\paragraph{Peak location variability.}
The peak location shifts between $N$ values (from
$\gamma \approx 0.62$ at $N=40$ to $\gamma \approx 1.08$
at $N=10$), consistent with the dynamical variability
discussed in the main text and not
indicative of a fixed critical point.

\begin{thebibliography}{10}
\expandafter\ifx\csname url\endcsname\relax
  \def\url#1{\texttt{#1}}\fi
\expandafter\ifx\csname urlprefix\endcsname\relax\def\urlprefix{URL }\fi
\expandafter\ifx\csname href\endcsname\relax
  \def\href#1#2{#2} \def\path#1{#1}\fi

\bibitem{ref11}
J.~Hofbauer, K.~Sigmund, Evolutionary Games and Population Dynamics, Cambridge
  University Press, Cambridge, 1998.
\newblock \href {https://doi.org/10.1017/CBO9781139173179}
  {\path{doi:10.1017/CBO9781139173179}}.

\bibitem{ref12}
A.~J. McKane, T.~J. Newman, Predator-prey cycles from resonant amplification of
  demographic stochasticity, Physical Review Letters 94~(21) (2005) 218102.
\newblock \href {https://doi.org/10.1103/PhysRevLett.94.218102}
  {\path{doi:10.1103/PhysRevLett.94.218102}}.

\bibitem{ref14}
C.~Taylor, M.~A. Nowak, Evolutionary game dynamics in finite populations,
  Bulletin of Mathematical Biology 66~(6) (2004) 1621--1644.
\newblock \href {https://doi.org/10.1016/j.bulm.2004.03.004}
  {\path{doi:10.1016/j.bulm.2004.03.004}}.

\bibitem{ref10}
W.~Horsthemke, R.~Lefever, Noise-Induced Transitions: Theory and Applications
  in Physics, Chemistry, and Biology, Springer, Berlin, 1984.
\newblock \href {https://doi.org/10.1007/3-540-36852-3}
  {\path{doi:10.1007/3-540-36852-3}}.

\bibitem{ref22}
C.~Van~den Broeck, J.~M.~R. Parrondo, R.~Toral, Noise-induced nonequilibrium
  phase transition, Physical Review Letters 73~(25) (1994) 3395--3398.
\newblock \href {https://doi.org/10.1103/PhysRevLett.73.3395}
  {\path{doi:10.1103/PhysRevLett.73.3395}}.

\bibitem{ref4}
M.~E. Fisher, M.~N. Barber, Scaling theory for finite-size effects in the
  critical region, Physical Review Letters 28~(23) (1972) 1516--1519.
\newblock \href {https://doi.org/10.1103/PhysRevLett.28.1516}
  {\path{doi:10.1103/PhysRevLett.28.1516}}.

\bibitem{ref5}
K.~Binder, Finite size scaling analysis of {I}sing model block distribution
  functions, Zeitschrift f{\"u}r Physik B 43~(2) (1981) 119--140.
\newblock \href {https://doi.org/10.1007/BF01293604}
  {\path{doi:10.1007/BF01293604}}.

\bibitem{ref7}
N.~G. van Kampen, Stochastic Processes in Physics and Chemistry, 3rd Edition,
  Elsevier, Amsterdam, 2007.
\newblock \href {https://doi.org/10.1016/B978-0-444-52965-7.X5000-4}
  {\path{doi:10.1016/B978-0-444-52965-7.X5000-4}}.

\bibitem{ref8}
C.~W. Gardiner, Stochastic Methods: A Handbook for the Natural and Social
  Sciences, 4th Edition, Springer, Berlin, 2009, iSBN: 978-3-540-70712-7.

\bibitem{ref17}
P.~E. Kloeden, E.~Platen, Numerical Solution of Stochastic Differential
  Equations, Springer, Berlin, 1992.
\newblock \href {https://doi.org/10.1007/978-3-662-12616-5}
  {\path{doi:10.1007/978-3-662-12616-5}}.

\bibitem{ref20}
D.~P. Landau, K.~Binder, A Guide to {M}onte {C}arlo Simulations in Statistical
  Physics, 5th Edition, Cambridge University Press, Cambridge, 2021.
\newblock \href {https://doi.org/10.1017/9781108780346}
  {\path{doi:10.1017/9781108780346}}.

\bibitem{ref26}
L.~A. Imhof, The long-run behavior of the stochastic replicator dynamics, Ann.
  Appl. Probab. 15~(1B) (2005) 1019--1045.
\newblock \href {https://doi.org/10.1214/105051604000000837}
  {\path{doi:10.1214/105051604000000837}}.

\bibitem{ref25}
S.~Bubeck, R.~Eldan, J.~Lehec, Sampling from a log-concave distribution with
  projected {L}angevin {M}onte {C}arlo, Discrete Comput. Geom. 59~(4) (2018)
  757--783.
\newblock \href {https://doi.org/10.1007/s00454-018-9992-1}
  {\path{doi:10.1007/s00454-018-9992-1}}.

\bibitem{ref24}
P.-L. Lions, A.-S. Sznitman, Stochastic differential equations with reflecting
  boundary conditions, Comm. Pure Appl. Math. 37 (1984) 511--537.
\newblock \href {https://doi.org/10.1002/cpa.3160370408}
  {\path{doi:10.1002/cpa.3160370408}}.

\bibitem{ref1}
H.~Hinrichsen, Non-equilibrium critical phenomena and phase transitions into
  absorbing states, Advances in Physics 49~(7) (2000) 815--958.
\newblock \href {https://doi.org/10.1080/00018730050198152}
  {\path{doi:10.1080/00018730050198152}}.

\bibitem{ref2}
P.~C. Hohenberg, B.~I. Halperin, Theory of dynamic critical phenomena, Reviews
  of Modern Physics 49~(3) (1977) 435--479.
\newblock \href {https://doi.org/10.1103/RevModPhys.49.435}
  {\path{doi:10.1103/RevModPhys.49.435}}.

\bibitem{ref3}
U.~C. T{\"a}uber, Critical Dynamics: A Field Theory Approach to Equilibrium and
  Non-Equilibrium Scaling Behavior, Cambridge University Press, Cambridge,
  2014.
\newblock \href {https://doi.org/10.1017/CBO9781139046213}
  {\path{doi:10.1017/CBO9781139046213}}.

\bibitem{ref21}
W.~Kob, H.~C. Andersen, Testing mode-coupling theory for a supercooled binary
  {L}ennard-{J}ones mixture, Physical Review E 51~(5) (1995) 4626--4641.
\newblock \href {https://doi.org/10.1103/PhysRevE.51.4626}
  {\path{doi:10.1103/PhysRevE.51.4626}}.

\bibitem{ref18}
T.~Biancalani, L.~Dyson, A.~J. McKane, Noise-induced bistable states and their
  mean switching time in foraging colonies, Physical Review Letters 112~(3)
  (2014) 038101.
\newblock \href {https://doi.org/10.1103/PhysRevLett.112.038101}
  {\path{doi:10.1103/PhysRevLett.112.038101}}.

\bibitem{ref13}
F.~Ritort, P.~Sollich, Glassy dynamics of kinetically constrained models,
  Advances in Physics 52~(4) (2003) 219--342.
\newblock \href {https://doi.org/10.1080/0001873031000093582}
  {\path{doi:10.1080/0001873031000093582}}.

\bibitem{ref15}
J.-P. Bouchaud, Weak ergodicity breaking and aging in disordered systems,
  Journal de Physique I 2~(9) (1992) 1705--1713.
\newblock \href {https://doi.org/10.1051/jp1:1992238}
  {\path{doi:10.1051/jp1:1992238}}.

\bibitem{ref16}
L.~Berthier, G.~Biroli, Theoretical perspective on the glass transition and
  amorphous materials, Reviews of Modern Physics 83~(2) (2011) 587--645.
\newblock \href {https://doi.org/10.1103/RevModPhys.83.587}
  {\path{doi:10.1103/RevModPhys.83.587}}.

\bibitem{ref23}
P.~Ashwin, S.~Wieczorek, R.~Vitolo, P.~Cox, Tipping points in open systems:
  bifurcation, noise-induced and rate-dependent examples in the climate system,
  Philosophical Transactions of the Royal Society A 370~(1962) (2012)
  1166--1184.
\newblock \href {https://doi.org/10.1098/rsta.2011.0306}
  {\path{doi:10.1098/rsta.2011.0306}}.

\bibitem{ref6}
V.~Privman (Ed.), Finite Size Scaling and Numerical Simulation of Statistical
  Systems, World Scientific, Singapore, 1990.
\newblock \href {https://doi.org/10.1142/1011} {\path{doi:10.1142/1011}}.

\bibitem{ref9}
H.~Risken, The {F}okker--{P}lanck Equation: Methods of Solution and
  Applications, 2nd Edition, Springer, Berlin, 1989.
\newblock \href {https://doi.org/10.1007/978-3-642-61544-3}
  {\path{doi:10.1007/978-3-642-61544-3}}.

\bibitem{ref19}
M.~Assaf, B.~Meerson, {WKB} theory of large deviations in stochastic
  populations, Journal of Physics A: Mathematical and Theoretical 50~(26)
  (2017) 263001.
\newblock \href {https://doi.org/10.1088/1751-8121/aa669a}
  {\path{doi:10.1088/1751-8121/aa669a}}.

\end{thebibliography}
\end{document}